\documentclass[onecolumn]{aa}
\usepackage{graphicx}
\usepackage{lscape}
\usepackage{rotate}
\usepackage{txfonts}
\def\lesssim{\mathrel{\hbox{\rlap{\hbox{\lower4pt\hbox{$\sim$}}}\hbox{$<$}}}}
\def\gtrsim{\mathrel{\hbox{\rlap{\hbox{\lower4pt\hbox{$\sim$}}}\hbox{$>$}}}}
\newcommand{\mincir}{\raise
-2.truept\hbox{\rlap{\hbox{$\sim$}}\raise5.truept
\hbox{$<$}\ }}
\newcommand{\magcir}{\raise
-2.truept\hbox{\rlap{\hbox{$\sim$}}\raise5.truept
\hbox{$>$}\ }}
\newcommand{\siml}{\raise -2.truept\hbox{\rlap{\hbox{$\sim$}}\raise5.truept
\hbox{$<$}\ }}
\newcommand{\simg}{\raise -2.truept\hbox{\rlap{\hbox{$\sim$}}\raise5.truept
\hbox{$>$}\ }}
\newcommand{\be}{\begin{equation}}
\newcommand{\ee}{\end{equation}}
\newcommand{\ba}{\begin{eqnarray}}
\newcommand{\ea}{\end{eqnarray}}

\newcommand {\kss} {km~s$^{-1}$}
\newcommand {\msun} {$\mathrm{h^{-1} \  M_{\odot} \;}$}

\begin{document}
   \title{Transformations of
galaxies in the environments of the cluster ABCG\,209 at z$\sim$0.2
\thanks{Based on observations collected at the European Southern
Observatory, Chile (Proposal ESO NM-0 68.AB-0116)}
}
   \titlerunning{Transformations of galaxies at z$\sim$0.2}

\author{
A. Mercurio\inst{1,2}
\and G. Busarello\inst{1}
\and P. Merluzzi \inst{1} 
\and F. La Barbera \inst{1}
\and M. Girardi \inst{2} 
\and C. P. Haines\inst{1}
}
\offprints{A. Mercurio}
\institute{INAF - Osservatorio Astronomico di Capodimonte,
via Moiariello 16, I-80131 Napoli, Italy\\
\and   Dipartimento di Astronomia, Universit\`{a} degli Studi di Trieste,
Via Tiepolo 11, I-34100 Trieste, Italy\\
}
\date{Received; accepted}

\abstract{We analyse the properties of galaxy populations in the rich
Abell cluster ABCG\,209 at redshift z$\sim$0.21, on the basis of
spectral classification of 102 member galaxies. We take advantage of
available structural parameters to study separately the properties of
bulge-dominated and disk-dominated galaxies.  The star formation
histories of the cluster galaxy populations are investigated by using
line strengths and the 4000 \AA \ \ break, through a comparison to
stellar population synthesis models. The dynamical properties of
different spectral classes are examined in order to infer the past
merging history of ABCG\,209. The cluster is characterized by the
presence of two components: an old galaxy population, formed very
early (z$_f \gtrsim $ 3.5), and a younger (z$_f \gtrsim $ 1.2)
population of infalling galaxies. We find evidence of a merger with an
infalling group of galaxies occurred 3.5-4.5 Gyr ago. The correlation
between the position of the young H$_\delta$-strong galaxies and the
X-ray flux shows that the hot intracluster medium triggered a
starburst in this galaxy population $\sim$ 3 Gyr ago.

 }

   \maketitle
%

\section{Introduction}

The influence of environment on the formation and the evolution of
galaxies remains one of the most pressing issues in cosmology. The
study of galaxy populations in rich clusters offers a unique
opportunity to observe directly galaxy evolution, and environmental
effects on star formation, by providing large numbers of galaxies at
the same redshift which have been exposed to a wide variety of
environments. In particular, galaxy populations in rich clusters are
different from those in poorer environments, suggesting that some
mechanism working on the cluster population is absent in the low
density or field environments.

The first evidence for significant evolution in the dense environments
of rich clusters over the past $\sim$ 5 Gyr was the discovery of the
Butcher-Oemler effect (Butcher \& Oemler \cite{but78}, \cite{but84}),
in which higher redshift clusters tend to have a larger fraction of
blue galaxies than those at the present epoch (but see also La Barbera
et al. \cite{lab03}). This purely photometric result was subsequently
confirmed through spectroscopic observations (e.g., Dressler \& Gunn
\cite{dre82}, \cite{dre83}, \cite{dre92}; Lavery \& Henry
\cite{lav86}; Couch \& Sharples \cite{cou87}), which have provided key
information concerning the nature of blue galaxies, showing that
several of them have strong emission line spectra, generally due to
the presence of active star formation. They have also shown the
existence of a class of galaxies, first identified by Dressler \& Gunn
(\cite{dre83}), which exhibit strong Balmer absorption lines without
detectable emissions. Dressler \& Gunn inferred the presence of a
substantial population of A--type stars in these galaxies and
concluded that a fraction of distant blue populations are
``post-starburst'' galaxies with abruptly truncated star
formation. In this scenario, however, the most surprising result is
that many red galaxies exhibit post-starburst spectra, indicative of a
recent enhancement of star formation (e.g. Couch \& Sharples
\cite{cou87}; Dressler \& Gunn \cite{dre92}).

Numerous studies have since been made to understand the evolution of
these galaxies and the connection with the galaxy populations observed
in clusters today (e.g. Couch \& Sharples \cite{cou87}; Barger et
al. \cite{bar96}; Couch et al. \cite{cou94}; Abraham et
al. \cite{abr96}; Morris et al. \cite{mor98}; Poggianti et
al. \cite{pog99}; Balogh et al. \cite{bal99}; Ellingson et
al. \cite{ell01}). The emerging picture is that of a galaxy population
formed early (z $\gtrsim$ 2) in the cluster's history, having
characteristically strong 4000 \AA \ \ breaks and red colours. Then
the clusters grow by infall of field galaxies. This accretion process
causes the truncation of star formation, possibly with an associated
starburst. As this transformation proceeds, these galaxies might be
identified with normal looking spirals, then as galaxies with strong
Balmer absorption spectra, and finally as S0 galaxies, which have
retained some of their disk structure but have ceased active star
formation (Dressler et al. \cite{dre97}).

On the other hand, there is observational evidence for a strong
connection between galaxy properties and the presence of
substructures, which indicates that cluster merger phenomena are still
ongoing. Caldwell et al. (\cite{cal93}) and Caldwell \& Rose
(\cite{cal97}) found that post-starburst galaxies are located
preferentially near a secondary peak in the X-ray emission of the Coma
cluster and suggested that merging between Coma and a group of
galaxies plays a vital role in triggering a secondary starburst in
galaxies (but see also Goto et al. \cite{got03a}, \cite{got03b}).

Numerical simulations have shown that cluster mergers may trigger
starbursts, and can affect greatly star formation histories. Tomita et
al. (\cite{tom96}) argued that in a merging cluster some galaxies may
experience a rapid increase of external pressure, passing through
regions which are overdense in intra-cluster gas, leading to a
starburst. An excess of star--forming galaxies is expected in the
regions between the colliding sub-clusters. As shown by Bekki
(\cite{bek99}), mergers induce a time--dependent tidal gravitational
field that stimulates non--axisymmetric perturbations in disk
galaxies, driving efficient transfer of gas to the central region, and
finally triggering a secondary starburst in the central part of these
galaxies. Roettiger et al. (\cite{roe96}) showed that during
cluster--cluster mergers a bow shock forms on the edges of infalling
subclusters, that protects the gas--rich subcluster galaxies from ram
pressure stripping. This protection fails at core crossing, and
galaxies initiate a burst of star formation.

The above results show that episodes of star formation in cluster
galaxies can be driven both by the accretion of field galaxies into
the cluster and by cluster-cluster merging. In order to disentangle
the two effects it is crucial to relate the star formation history of
cluster galaxies to global properties of clusters, such as mass and
dynamical state.

To examine the effect of environment and dynamics on galaxy
properties, in particular on star formation, we have performed a
spectroscopic investigation of luminous galaxies in the galaxy cluster
ABCG\,209 at z=0.21 (Mercurio et al.  \cite{mer03a} and references
therein) using EMMI-NTT spectra. ABCG\,209 is a rich (richness class
R=3, Abell et al. \cite{abe89}) , X-ray luminous
(L$_{\mathrm{X}}$=(0.1--2.4\,keV)$\sim2.7\times10^{45}\,h^{-2}_{70}\,$erg
s$^{-1}$, Ebeling et al. \cite{ebe96}; T$_{\mathrm{X}}\sim10$ keV,
Rizza et al. \cite{riz98}), and massive cluster
($\mathrm{M(<R_{vir})=2.25^{+0.63}_{-0.65}\times10^{15}}$ \msun,
Mercurio et al.  \cite{mer03a}). It is characterized by the presence
of substructure, allowing the effect of cluster dynamics and evolution
on the properties of its member galaxies to be examined. Evidence in
favour of the cluster undergoing strong dynamical evolution is found
in the form of a velocity gradient along a SE-NW axis, which is the
same preferential direction found from the elongation in the spatial
distribution of galaxies and of the X-ray flux as well as that of the
cD galaxy, and in the presence of substructures. These substructures
are manifest both in the elongation and asymmetry of the X-ray
emission, with two main clumps (Rizza et al. \cite{riz98}) and from
the analysis of the velocity distribution (Mercurio et
al. \cite{mer03a}). Moreover, the young dynamical state of the cluster
is indicated by the possible presence of a radio halo (Giovannini,
Tordi \& Feretti \cite{gio99}), which has been suggested to be the
result of a recent cluster merger, through the acceleration of
relativistic particles by the merger shocks (Feretti \cite{fer02}).

The data are presented in Sect.~\ref{sec:2}, and the spectroscopic
measurement in Sect.~\ref{sec:3}. In Sect.~\ref{sec:4} we discuss the
spectral classification comparing data with models with different
metallicities and star formation histories. We study separately the
properties of disks and spheroids, on the basis of the Sersic index,
in Sect.~\ref{sec:5}. We derive the age distribution of different
spectral classes in Sect.~\ref{sec:6}. The analysis of the velocity
distribution and of possible segregation is presented in
Sect.~\ref{sec:7}. Finally we summarize and discuss the results in
Sect.~\ref{sec:8}.

Throughout the paper, we use the convention that equivalent widths are
positive for absorption lines and negative for emission lines.

We adopt a flat cosmology with $\mathrm{\Omega_M=0.3}$,
$\mathrm{\Omega_{\Lambda}=0.7}$, and H$_0$ = 70 $h_{70}$ km s$^{-1}$
Mpc$^{-1}$. With this cosmological model, the age of the universe is
13.5 Gyr, and the look-back time at z=0.21 is 2.5 Gyr.


\section{The data}
\label{sec:2}

The observations of the galaxy cluster ABCG\,209 consist of
spectroscopic and photometric data, collected at the ESO New
Technology Telescope (NTT) with the ESO Multi Mode Instrument (EMMI)
in October 2001, and archive Canada-France-Hawaii Telescope (CFHT)
images.

Spectroscopic data have been obtained in the multi--object
spectroscopy (MOS) mode of EMMI. Targets were randomly selected by
using preliminary R--band images (T$\mathrm{_{exp}}$=180 s) to
construct the multislit plates. A total of 112 cluster members with R
$\lesssim$ 20.0 were observed in four fields (field of view $5^\prime
\times 8.6^\prime$), with different position angles on the sky.  We
exposed the masks with integration times from 0.75 to 3 hr with the
EMMI--Grism\#2, yielding a dispersion of $\sim2.8$ \AA/pix and a
resolution of $\sim 8$ \AA \ \ FWHM, in the spectral range 385 -- 900
nm.  Reduction procedures used for the spectroscopic data are
described in Mercurio et al. (\cite{mer03a}). In order to perform the
flux calibration, the spectrophotometric standard star LTT 7987 from
Hamuy et al. (\cite{ham94}) was also observed.

The photometric data were collected on November 1999 (PI. J.-P. Kneib)
using the CFH12K mosaic camera, consisting of B-- and R--band wide
field images, centred on the cluster ABCG\,209 and covering a total
field of view of $42^\prime \times 28^\prime$ (8.6 $\times$ 5.7
$\mathrm{h^{-2}_{70}}$ Mpc$^2$ at the cluster redshift). The CCDs have
a pixel scale of 0.206 arcsec. The total exposure times for both B and
R images were 7200s and the seeing was 1.02$^{\prime\prime}$ in B and
0.73$^{\prime\prime}$ in R.  Reduction procedures, photometric
calibration and catalogue extraction are described in Haines et
al. (\cite{hai03}).

We classified each member galaxy on the basis of the measured
equivalent widths of the atomic features [OII], H$_{\delta A}$, [OIII]
, H$_{\alpha}$ and of the strength of the 4000 \AA \ \ break (see
Sect. \ref{sec:3} and Sect. \ref{sec:4}). Out of 112 cluster members,
we were not able to obtain the spectral classification for 10
galaxies, because the spectra covered a too short wavelength range on
the CCD.  We also studied the properties of bulge-dominated and
disk-dominated galaxies separately, based on the value of the Sersic
index $n$ in the R--band. The Sersic index $n$ was derived in La
Barbera et al. (\cite{lab02}, \cite{lab03}). Table \ref{phot_cat}
presents the photometric properties of the member galaxies.

\begin{table}
        \caption[]{\footnotesize Photometric data. Running number for
         galaxies in the present sample, ID (Col.~1); right ascension 
	 and declination (Col.~2, Col.~3) R-band Kron
         magnitude (Col.~4); B-R colour (Col.~5); heliocentric
         corrected redshift $\mathrm{z}$ (Col.~6); Sersic index $n$
         (Col.~7). Errors in magnitudes and colours are rounded to
         the second decimal place.}
         \label{phot_cat}
{\scriptsize
         $$ 
           \begin{array}{c c c c c c c|c c c c c c c}
            \hline
            \noalign{\smallskip}
            \hline
            \noalign{\smallskip}
\mathrm{ID} & \mathrm{RA(J2000)} & \mathrm{DEC(J2000)} & \mathrm{R} & 
\mathrm{B-R} & \mathrm{z} & n &\mathrm{ID} & \mathrm{RA(J2000)} &
\mathrm{DEC(J2000)} & \mathrm{R} & \mathrm{B-R} & \mathrm{z} & n\\
            \hline
            \noalign{\smallskip}   
  1& 01\ 31\ 33.81 & -13\ 32\ 22.9 & 17.77\pm0.01& 1.69\pm0.01& 0.2191\pm0.0003& 2.8& 52& 01\ 31\ 51.73 & -13\ 38\ 40.2 & 18.46\pm0.01& 2.37\pm0.01& 0.2184\pm0.0002& 5.4\\
  2& 01\ 31\ 33.82 & -13\ 38\ 28.5 & 18.49\pm0.01& 2.40\pm0.01& 0.2075\pm0.0002& 2.3& 53& 01\ 31\ 51.82 & -13\ 38\ 30.2 & 19.44\pm0.01& 2.33\pm0.02& 0.2028\pm0.0005& 3.8\\
  3& 01\ 31\ 33.86 & -13\ 32\ 43.4 & 18.54\pm0.01& 2.37\pm0.01& 0.1998\pm0.0003& 2.8& 54& 01\ 31\ 52.31 & -13\ 36\ 57.9 & 17.19\pm0.01& 2.35\pm0.01& 0.2024\pm0.0002& 5.3\\
  4& 01\ 31\ 34.11 & -13\ 32\ 26.8 & 18.99\pm0.01& 1.82\pm0.01& 0.2000\pm0.0004& 3.6& 55& 01\ 31\ 52.54 & -13\ 36\ 40.4 & 16.41\pm0.01& 2.54\pm0.01& 0.2097\pm0.0002& 6.6\\
  5& 01\ 31\ 34.26 & -13\ 38\ 13.1 & 19.42\pm0.01& 2.30\pm0.02& 0.2145\pm0.0005& 1.6& 56& 01\ 31\ 53.34 & -13\ 36\ 31.3 & 18.46\pm0.01& 2.28\pm0.01& 0.2094\pm0.0002& 2.9\\
  6& 01\ 31\ 35.37 & -13\ 31\ 18.9 & 16.62\pm0.01& 2.43\pm0.01& 0.2068\pm0.0004& 3.2& 57& 01\ 31\ 53.86 & -13\ 38\ 21.2 & 19.00\pm0.01& 1.85\pm0.01& 0.2170\pm0.0003& 1.4\\
  7& 01\ 31\ 35.37 & -13\ 37\ 17.4 & 17.74\pm0.01& 2.49\pm0.01& 0.2087\pm0.0002& 3.1& 58& 01\ 31\ 53.87 & -13\ 36\ 13.4 & 18.11\pm0.01& 2.42\pm0.01& 0.2085\pm0.0002& 2.5\\
  8& 01\ 31\ 35.42 & -13\ 34\ 52.0 & 18.21\pm0.01& 2.37\pm0.01& 0.2073\pm0.0002& 8.4& 59& 01\ 31\ 54.33 & -13\ 38\ 54.9 & 19.29\pm0.01& 2.23\pm0.01& 0.2083\pm0.0004& 4.81\\
  9& 01\ 31\ 35.61 & -13\ 32\ 51.3 & 19.28\pm0.01& 2.41\pm0.01& 0.2090\pm0.0002& 4.5& 60& 01\ 31\ 55.12 & -13\ 37\ 04.4 & 18.58\pm0.01& 2.44\pm0.01& 0.2118\pm0.0003& 7.3\\
 10& 01\ 31\ 36.51 & -13\ 33\ 45.2 & 19.07\pm0.01& 2.33\pm0.01& 0.2072\pm0.0004& 2.4& 61& 01\ 31\ 55.18 & -13\ 36\ 57.6 & 18.40\pm0.01& 2.41\pm0.01& 0.2150\pm0.0002& 8.4\\
 11& 01\ 31\ 36.87 & -13\ 39\ 19.3 & 18.95\pm0.01& 2.38\pm0.01& 0.2039\pm0.0002& 4.6& 62& 01\ 31\ 48.20 & -13\ 38\ 12.3 & 18.82\pm0.01& 0.95\pm0.01& 0.2188\pm0.0001& 1.2\\
 12& 01\ 31\ 37.23 & -13\ 32\ 09.9 & 19.17\pm0.01& 2.37\pm0.01& 0.2084\pm0.0003& 4.2& 63& 01\ 31\ 55.47 & -13\ 38\ 28.9 & 18.34\pm0.01& 2.35\pm0.01& 0.2144\pm0.0002& 4.8\\
 13& 01\ 31\ 37.38 & -13\ 34\ 46.3 & 17.90\pm0.01& 2.36\pm0.01& 0.2134\pm0.0003& 4.2& 64& 01\ 31\ 55.95 & -13\ 36\ 40.4 & 17.66\pm0.01& 1.43\pm0.01& 0.1999\pm0.0005& 0.8\\
 14& 01\ 31\ 38.06 & -13\ 33\ 35.9 & 19.06\pm0.01& 2.48\pm0.01& 0.2073\pm0.0002& 1.8& 65& 01\ 31\ 56.22 & -13\ 36\ 46.7 & 18.10\pm0.01& 2.11\pm0.01& 0.2098\pm0.0003& 1.4\\
 15& 01\ 31\ 38.69 & -13\ 35\ 54.7 & 19.40\pm0.01& 2.20\pm0.02& 0.2084\pm0.0003& 3.1& 66& 01\ 31\ 56.58 & -13\ 38\ 34.9 & 19.28\pm0.01& 2.35\pm0.02& 0.2117\pm0.0002& 4.8\\
 16& 01\ 31\ 39.06 & -13\ 33\ 39.9 & 18.85\pm0.01& 2.32\pm0.01& 0.2120\pm0.0002& 3.9& 67& 01\ 31\ 56.78 & -13\ 40\ 04.8 & 19.40\pm0.01& 2.29\pm0.02& 0.2098\pm0.0004& 4.0\\
 17& 01\ 31\ 39.89 & -13\ 35\ 45.1 & 18.35\pm0.01& 1.66\pm0.01& 0.2073\pm0.0002& 1.5& 68& 01\ 31\ 56.91 & -13\ 38\ 30.2 & 17.46\pm0.01& 2.44\pm0.01& 0.2102\pm0.0002& 4.9\\
 18& 01\ 31\ 40.05 & -13\ 32\ 08.9 & 18.67\pm0.01& 2.33\pm0.01& 0.2096\pm0.0003& 4.6& 69& 01\ 31\ 57.38 & -13\ 38\ 08.2 & 18.76\pm0.01& 2.43\pm0.01& 0.2107\pm0.0003& 5.1\\
 19& 01\ 31\ 40.17 & -13\ 36\ 06.2 & 18.46\pm0.01& 2.18\pm0.01& 0.1995\pm0.0002& 5.2& 70& 01\ 31\ 57.99 & -13\ 38\ 59.7 & 18.33\pm0.01& 2.38\pm0.01& 0.1973\pm0.0003& 2.8\\
 20& 01\ 31\ 40.58 & -13\ 36\ 32.9 & 18.61\pm0.01& 2.19\pm0.01& 0.2052\pm0.0003& 1.6& 71& 01\ 31\ 58.28 & -13\ 39\ 36.2 & 18.34\pm0.01& 2.39\pm0.01& 0.2056\pm0.0002& 2.9\\
 21& 01\ 31\ 40.76 & -13\ 34\ 17.0 & 17.83\pm0.01& 2.33\pm0.01& 0.2051\pm0.0002& 3.0& 72& 01\ 31\ 58.68 & -13\ 38\ 04.0 & 19.61\pm0.01& 2.20\pm0.02& 0.2093\pm0.0004& 7.7\\
 22& 01\ 31\ 40.94 & -13\ 37\ 36.5 & 18.53\pm0.01& 1.66\pm0.01& 0.2181\pm0.0008& 3.2& 73& 01\ 31\ 59.10 & -13\ 39\ 26.4 & 18.17\pm0.01& 2.37\pm0.01& 0.2104\pm0.0002& 8.4\\
 23& 01\ 31\ 41.63 & -13\ 37\ 32.3 & 18.60\pm0.01& 2.28\pm0.01& 0.2004\pm0.0002& 5.6& 74& 01\ 32\ 00.34 & -13\ 37\ 56.5 & 19.30\pm0.01& 2.30\pm0.01& 0.2056\pm0.0002& 2.3\\
 24& 01\ 31\ 41.64 & -13\ 38\ 50.2 & 18.81\pm0.01& 1.58\pm0.01& 0.2063\pm0.0003& 2.3& 75& 01\ 32\ 01.18 & -13\ 35\ 17.5 & 18.87\pm0.01& 2.32\pm0.01& 0.2172\pm0.0003& 5.5\\
 25& 01\ 31\ 42.35 & -13\ 39\ 26.0 & 18.28\pm0.01& 2.28\pm0.01& 0.2051\pm0.0004& 7.7& 76& 01\ 32\ 01.66 & -13\ 35\ 33.8 & 17.47\pm0.01& 2.45\pm0.01& 0.2069\pm0.0002& 3.0\\
 26& 01\ 31\ 42.76 & -13\ 38\ 31.6 & 18.95\pm0.01& 2.30\pm0.01& 0.2034\pm0.0002& 5.0& 77& 01\ 32\ 01.82 & -13\ 33\ 34.0 & 17.75\pm0.01& 2.38\pm0.01& 0.2083\pm0.0003& 5.0\\
 27& 01\ 31\ 43.69 & -13\ 35\ 57.8 & 20.00\pm0.01& 2.34\pm0.02& 0.2076\pm0.0003& 7.9& 78& 01\ 32\ 01.84 & -13\ 36\ 15.7 & 19.49\pm0.01& 2.16\pm0.02& 0.2131\pm0.0004& 2.2\\
 28& 01\ 31\ 43.81 & -13\ 39\ 27.7 & 19.66\pm0.01& 2.16\pm0.02& 0.2064\pm0.0002& 4.0& 79& 01\ 32\ 01.91 & -13\ 35\ 31.2 & 18.04\pm0.01& 1.91\pm0.01& 0.1984\pm0.0002& 1.9\\
 29& 01\ 31\ 44.47 & -13\ 37\ 03.2 & 19.32\pm0.01& 2.02\pm0.01& 0.2027\pm0.0003& 0.8& 80& 01\ 32\ 02.18 & -13\ 35\ 51.0 & 18.40\pm0.01& 2.10\pm0.01& 0.2091\pm0.0003& 2.6\\
 30& 01\ 31\ 45.24 & -13\ 37\ 39.7 & 17.64\pm0.01& 2.25\pm0.01& 0.2060\pm0.0002& 3.6& 81& 01\ 32\ 02.47 & -13\ 32\ 13.0 & 19.31\pm0.01& 1.76\pm0.01& 0.1982\pm0.0004& 0.8\\
 31& 01\ 31\ 45.61 & -13\ 39\ 02.0 & 19.11\pm0.01& 2.36\pm0.01& 0.2087\pm0.0004& 3.1& 82& 01\ 32\ 02.94 & -13\ 39\ 39.4 & 18.71\pm0.01& 2.46\pm0.01& 0.2110\pm0.0002& 3.2\\
 32& 01\ 31\ 45.87 & -13\ 36\ 38.0 & 18.07\pm0.01& 2.37\pm0.01& 0.2066\pm0.0002& 2.8& 83& 01\ 32\ 03.07 & -13\ 36\ 44.0 & 19.19\pm0.01& 2.21\pm0.02& 0.2102\pm0.0005& 5.6\\
 33& 01\ 31\ 46.15 & -13\ 34\ 56.6 & 17.75\pm0.01& 2.43\pm0.01& 0.2084\pm0.0002& 2.3& 84& 01\ 32\ 03.50 & -13\ 31\ 59.1 & 19.04\pm0.01& 1.74\pm0.01& 0.1971\pm0.0003& 1.1\\
 34& 01\ 31\ 46.33 & -13\ 37\ 24.5 & 19.53\pm0.01& 2.23\pm0.02& 0.2105\pm0.0003& 4.1& 85& 01\ 32\ 03.96 & -13\ 35\ 54.4 & 19.13\pm0.01& 2.32\pm0.01& 0.2137\pm0.0003& 2.0\\
 35& 01\ 31\ 46.58 & -13\ 38\ 40.9 & 18.24\pm0.01& 2.32\pm0.01& 0.2167\pm0.0002& 4.1& 86& 01\ 32\ 04.29 & -13\ 39\ 53.9 & 17.20\pm0.01& 2.44\pm0.01& 0.2125\pm0.0002& 9.2\\
 36& 01\ 31\ 47.26 & -13\ 33\ 10.3 & 18.88\pm0.01& 2.24\pm0.01& 0.2038\pm0.0004& 3.9& 87& 01\ 32\ 04.35 & -13\ 37\ 26.3 & 17.56\pm0.01& 2.47\pm0.01& 0.2061\pm0.0002& 5.0\\
 37& 01\ 31\ 47.69 & -13\ 37\ 50.4 & 17.34\pm0.01& 2.47\pm0.01& 0.2125\pm0.0003& 6.5& 88& 01\ 32\ 05.02 & -13\ 33\ 42.0 & 19.61\pm0.01& 2.22\pm0.02& 0.2087\pm0.0003& 2.7\\
 38& 01\ 31\ 47.91 & -13\ 39\ 07.9 & 17.76\pm0.01& 2.47\pm0.01& 0.2097\pm0.0003& 3.9& 89& 01\ 32\ 05.05 & -13\ 37\ 33.3 & 18.56\pm0.01& 2.33\pm0.01& 0.2074\pm0.0002& 7.5\\
 39& 01\ 31\ 48.01 & -13\ 38\ 26.7 & 18.54\pm0.01& 2.39\pm0.01& 0.2161\pm0.0002& 1.9& 90& 01\ 32\ 07.98 & -13\ 41\ 31.4 & 19.16\pm0.01& 2.23\pm0.01& 0.1997\pm0.0002& 2.2\\
 40& 01\ 31\ 48.45 & -13\ 37\ 43.2 & 19.64\pm0.01& 2.21\pm0.02& 0.1979\pm0.0005& 2.3& 91& 01\ 32\ 10.37 & -13\ 37\ 24.3 & 18.74\pm0.01& 2.39\pm0.01& 0.2161\pm0.0002& 2.9\\
 41& 01\ 31\ 49.21 & -13\ 37\ 35.0 & 18.88\pm0.01& 2.21\pm0.01& 0.2039\pm0.0003& 3.3& 92& 01\ 32\ 12.34 & -13\ 34\ 21.1 & 19.10\pm0.01& 1.72\pm0.01& 0.2134\pm0.0004& 0.7\\
 42& 01\ 31\ 49.38 & -13\ 36\ 06.9 & 19.76\pm0.01& 2.29\pm0.02& 0.2133\pm0.0004& 2.7& 93& 01\ 32\ 14.04 & -13\ 38\ 08.5 &  17.21\pm0.01& 2.44\pm0.01& 0.2175\pm0.0002& 4.\\
 43& 01\ 31\ 49.47 & -13\ 37\ 26.5 & 17.57\pm0.01& 1.55\pm0.01& 0.2140\pm0.0002& 1.4& 94& 01\ 32\ 12.51 & -13\ 41\ 18.1 & 18.70\pm0.01& 2.03\pm0.01& 0.2145\pm0.0002& 1.4\\
 44& 01\ 31\ 49.64 & -13\ 35\ 21.6 & 18.94\pm0.01& 2.42\pm0.01& 0.2061\pm0.0002& 4.0& 95& 01\ 32\ 15.00 & -13\ 41\ 13.9 & 17.96\pm0.01& 2.37\pm0.01& 0.2158\pm0.0003& 2.9\\
 45& 01\ 31\ 49.84 & -13\ 36\ 11.7 & 19.41\pm0.01& 2.05\pm0.02& 0.2123\pm0.0003& 3.0& 96& 01\ 32\ 15.56 & -13\ 37\ 49.2 & 18.15\pm0.01& 2.40\pm0.01& 0.2149\pm0.0002& 2.8\\
 46& 01\ 31\ 50.64 & -13\ 33\ 36.4 & 17.28\pm0.01& 2.43\pm0.01& 0.2064\pm0.0004& 5.6& 97& 01\ 32\ 15.84 & -13\ 35\ 41.0 & 19.98\pm0.01& 2.05\pm0.02& 0.2124\pm0.0004& 2.7\\
 47& 01\ 31\ 50.89 & -13\ 36\ 04.2 & 18.03\pm0.01& 2.43\pm0.01& 0.2078\pm0.0002& 3.4& 98& 01\ 32\ 16.00 & -13\ 38\ 34.6 & 18.82\pm0.01& 2.39\pm0.01& 0.2203\pm0.0003& 2.8\\
 48& 01\ 31\ 50.98 & -13\ 36\ 49.5 & 19.06\pm0.01& 2.27\pm0.02& 0.2042\pm0.0003& 3.4& 99& 01\ 32\ 16.20 & -13\ 32\ 38.8 & 18.19\pm0.01& 2.24\pm0.01& 0.2034\pm0.0003& 3.7\\
 49& 01\ 31\ 51.15 & -13\ 38\ 12.8 & 17.55\pm0.01& 2.31\pm0.01& 0.2183\pm0.0002& 6.3&100& 01\ 32\ 16.76 & -13\ 35\ 03.7 & 20.11\pm0.01& 1.98\pm0.02& 0.2138\pm0.0004& 1.2\\
 50& 01\ 31\ 51.34 & -13\ 36\ 56.6 & 17.49\pm0.01& 2.40\pm0.01& 0.2068\pm0.0002& 3.7&101& 01\ 32\ 16.98 & -13\ 33\ 01.1 & 19.88\pm0.01& 2.23\pm0.02& 0.2117\pm0.0004& 3.2\\
 51& 01\ 31\ 51.58 & -13\ 35\ 07.0 & 17.55\pm0.01& 2.16\pm0.01& 0.2001\pm0.0002& 4.0&102& 01\ 32\ 17.48 & -13\ 38\ 42.3 & 19.46\pm0.01& 1.28\pm0.01& 0.1963\pm0.0006& 1.0\\
        \noalign{\smallskip}	     		    
            \hline			    		    
        \noalign{\smallskip}	     		    
            \hline			    		    
         \end{array}
     $$ 
}
         \end{table}

\section{Derivation of line indices}
\label{sec:3}

In order to measure line indices, galaxy spectra were corrected for
galactic extinction following Schlegel et al. (\cite{sch98}),
flux--calibrated and wavelength--calibrated by using the
IRAF~\footnote{IRAF is distributed by the National Optical Astronomy
Observatories, which are operated by the Association of Universities
for Research in Astronomy, Inc., under cooperative agreement with the
National Science Foundation.} package ONEDSPEC.

The flux calibration was performed using observations of the
spectrophotometric standard LTT 7987, whose spectrum was acquired
before and after each scientific exposure.  The sensitivity function,
derived by using the tasks STANDARD and SENSFUNC (rms $\sim$ 0.03),
was applied to the galaxy spectra by using the task CALIBRATE. Within
this task, the exposures are corrected for the atmospheric extinction,
divided by the exposure time, and finally transformed using the
sensitivity curve. The calibrated spectra were then corrected for the
measured velocity dispersion by using DISPCOR.

In order to asses the quality of the flux calibration, we compared the
V-R colours as derived from photometry (Mercurio et al. \cite{mer03b})
with those measured from the spectra of 41 galaxies for which the
observed wavelength range is greater than 4750--8750 \AA. The
resulting mean difference in colours turns out to be $\Delta$(V-R) =
0.06$\pm$0.08, proving that the derived fluxes are correct.

We measured the equivalent widths of the following atomic and
molecular features: H$_{\delta A}$, H$_{\gamma A}$, Fe4531, H$_\beta$,
Fe5015, Mg$_1$, Mg$_2$, Mg$b$, Fe5270, Fe5335. We adopted the
definition of the extended Lick system (Worthey et al. \cite{wor94},
Worthey \& Ottaviani \cite{wor97}; Trager et al. \cite{tra98}), where
the spectral indices are determined in terms of a central feature
bandpass bracketed by two pseudo-continuum bandpasses at a resolution
of $\sim$9 \AA \ \ FWHM, that is similar to
ours\footnote{\footnotesize The difference of 1 \AA \ \ in resolution
between our data and the extended Lick standard results in a
difference in line indices amounting to about 1\% of the uncertainty
of the measurements.}.  Following the convention, atomic indices are
expressed in Angstroms of equivalent width, while molecular indices
are expressed in magnitudes.

We also measured the equivalent widths for the emission lines [OII],
[OIII] and H$_{\alpha}$ (see Table \ref{indices} for the definition of
wavelength ranges), as well as the strength of the 4000 \AA \ \ break
D$_n$(4000). We adopted the definition of D$_n$(4000) by Balogh et
al. (\cite{bal99}) as the ratio of the average flux in the narrow
bands 3850--3950 and 4000--4100 \AA. The original definition of this
index by Bruzual (\cite{bru83}) uses wider bands
(3750--3950 and 4050--4250 \AA), and hence it is more sensitive to
reddening effects.

\begin{table}
     \caption[]{Wavelength ranges for equivalent widths of emission lines, 
                expressed in Angstrom.} 
    \label{indices}
    $$
           \begin{array}{c c c c}
            \hline
            \noalign{\smallskip}
    \mathrm{Name} & \mathrm{Index \ bandpass} & \mathrm{Blue \ continuum} & \mathrm{Red \ continuum}\\
            \noalign{\smallskip}
            \hline
            \noalign{\smallskip}
\mathrm{[OII]}      & 3713-3741 & 3653-3713 & 3741-3801\\
\mathrm{[OIII]}     & 4997-5017 & 4872-4932 & 5050-5120\\
\mathrm{H_{\alpha}} & 6555-6575 & 6510-6540 & 6590-6620\\
            \noalign{\smallskip}
            \hline
         \end{array}
     $$
   \end{table}

The derivation of equivalent widths and the catalogue of spectroscopic
measurements are presented in Appendix A.

\section{Spectral classification}
\label{sec:4}

\begin{figure*}
  \centering 
\includegraphics[width=0.7\textwidth]{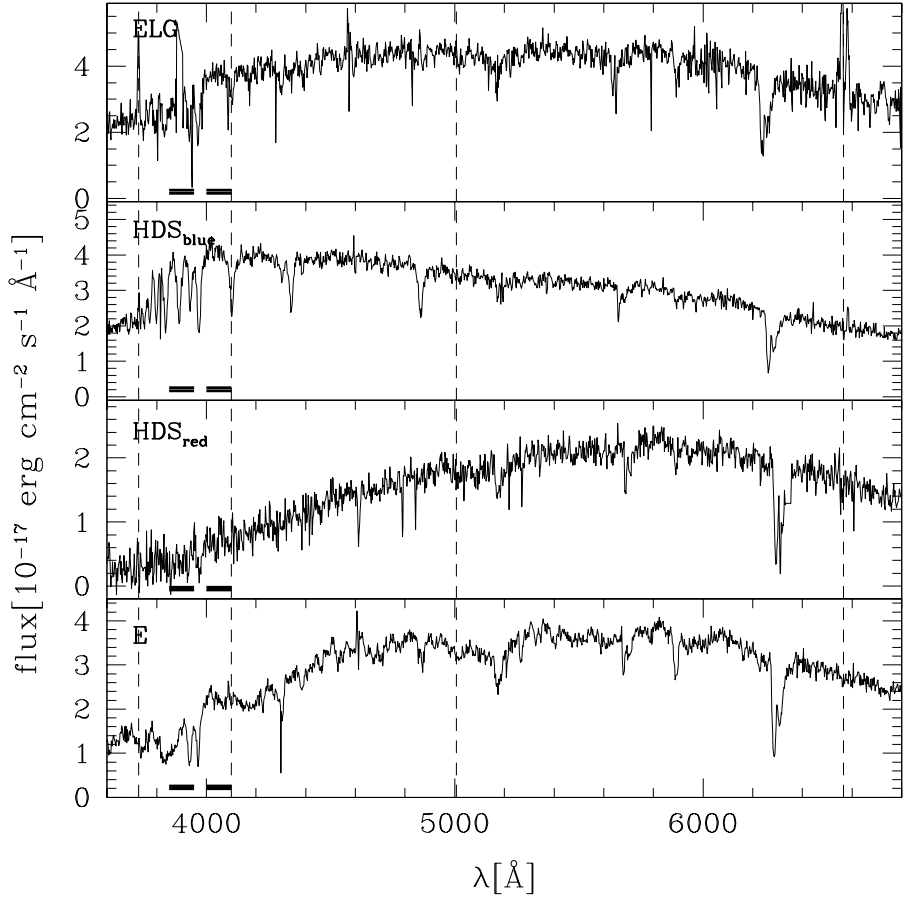} 
\caption{Rest-frame spectra of galaxies belonging to different
spectral classes. Lines [OII], H$_{\delta}$, [OIII],
H$_{\alpha}$ are highlighted by dashed lines. Wavelength ranges
adopted to measure D$_n$(4000) are also indicated by black lines.}
  \label{spectra}
\end{figure*}

An useful tool to classify galaxies is the diagram of D$_n$(4000)
versus H$_\delta$ equivalent width (EW(H$_\delta$)) (or
(B-R)-EW(H$_\delta$)); e.g. Couch \& Sharples \cite{cou87}, Barger et
al. \cite{bar96}) and the emission line measurements (e.g. [OII]). In
this way galaxies can be divided in four classes: emission line (ELG),
blue H$_{\delta}$ strong (HDS$_{\mathrm{blue}}$), red H$_{\delta}$
strong (HDS$_{\mathrm{red}}$) and passive (P) galaxies (see
Fig.~\ref{spectra}).

We remark that there is not a unique criterion in literature for the
definition of spectral classes of galaxies, mainly because there is no
general agreement on the definition of indices, and because the
spectra vary in terms of S/N and resolution. Moreover, the comparison
of spectral indices among different works is not straightforward since
line indices are sensitive to the definition of the bands and to the
method used to measure their strengths. Therefore, we derived spectral
indices for both data and models with the same procedures and we used
models themselves to define the spectral classes (see below).

\subsection{Galaxy population models}

To investigate the nature of galaxy populations in ABCG\,209, we used
the code GISSEL03 (Bruzual \& Charlot \cite{bru03}), by adopting a
Salpeter (\cite{sal55}) initial mass function (IMF), with a stellar
mass range from 0.1 to 100 $\mathrm{M_{\odot}}$. Results vary only
slightly choosing different IMFs (but maintaining the same mass
range).

The GISSEL03 code allows the spectral evolution of stellar populations
to be computed at a resolution of 3 \AA (FWHM), with a sampling of 1
\AA \ \ across the wavelength range 3200-9500 \AA. Differently from
previous models (Bruzual \& Charlot \cite{bru93}), this code provides
galaxy spectra with different metallicities, in the range
Z=(0.005--2.5)Z$_\odot$, which are required to break the
age-metallicity degeneracy, through a comparison of age and
metallicity sensitive indices (see Bruzual \& Charlot \cite{bru03} for
details).

Models were first degraded and resampled in order to match data and
then D$_n$(4000) and rest-frame equivalent widths were computed in the
same manner as for observations. The free parameters in our analysis
are the star formation rate and the age of the galaxies, while the
adopted metallicities were Z=0.4Z$_\odot$, Z$_\odot$, and
2.5Z$_\odot$.

\subsection{Emission line galaxies}

We define emission line galaxies as those showing in the spectra the
emission lines [OII] and H$_{\alpha}$ significant at $\ge$ 1 $\sigma$,
and [OIII]. When the [OII] line is out of the observed wavelength
range, we require [OIII] and H$_{\alpha}$ emission lines with $\ge$ 1
$\sigma$ to define an emission line galaxy (see Table
\ref{catalogue}). We decided to use simultaneously these three lines
and not only [OII], because i) we lack information on [OII] for more
than half of spectra (because of the available wavelength range), ii)
[OII] is heavily obscured by dust, and iii) [OII] is found at
wavelengths where the noise is higher.

This class contains star-forming (SF) and short starburst (SSB)
galaxies (Balogh et al. \cite{bal99}). We interpret these as systems
undergoing a star formation, that can be described by a model with
exponentially decaying SFR model with decay parameter
$\mu$=0.01\footnote{\footnotesize $\mu$=1-exp(1.0 Gyr/$\tau$)} (Barger
at al. \cite{bar96}). SSB galaxies, on the contrary, have experienced
a large increase of star formation over a short time, and can be
described by models with an initial burst of 100 Myr.

We detect emission lines in $\sim$ 7\% of cluster members.

\subsection{H$_{\delta}$-strong galaxies}

\begin{figure*}
  \centering 
\includegraphics[width=0.7\textwidth,bb= 10 330 377 698,clip]{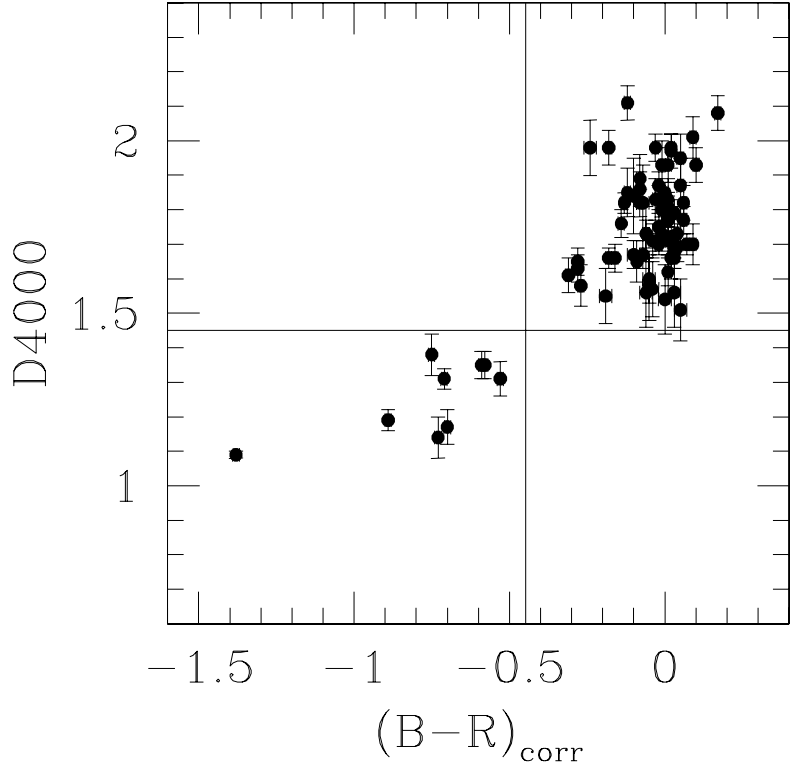} 
\vspace{-1.0cm} 
\caption{Observed distribution of D$_{n}$(4000) versus
(B--R)$_{\mathrm{corr}}$ (see text) for 76 member galaxies for which
we can measure D$_{n}$(4000). The vertical line indicates the
separation between blue and red galaxies as defined in Haines et
al. (\cite{hai03}) and the horizontal line is the cut for the break at
4000 \AA, adopted by Balogh et al. (\cite{bal99}).}
  \label{BR_D4000}
\end{figure*}

In order to investigate the physical properties of HDS galaxies, we
divided the sample into blue and red galaxies, according to B-R
colours corrected for the colour-magnitude (CM) relation:
(B-R)$_{\mathrm{corr}}$= (B-R)-(3.867 - 0.0815$\cdot$R). This relation
is obtained by performing MonteCarlo realisations of the cluster
populations, and fitting the resulting photometric data of $\sim$ 480
R $<$ 21.0 galaxies within one virial radius with the biweight
algorithm (see Haines et al. \cite{hai03} for more details).

We define as ``blue'' the galaxies with rest-frame B-V colours at
least 0.2\,mag bluer than that of the CM relation, as for the original
studies of Butcher \& Oemler (\cite{but84}). We estimate the
corresponding change in B-R colour at the cluster redshift by
firstly considering two model galaxies at $z=0$ (Bruzual \& Charlot
\cite{bru03}), one chosen to be a typical early-type galaxy 10\,Gyr
old with solar metallicity, and the second reduced in age until its
B-V colour becomes 0.2\,magnitude bluer. After moving both galaxies
to the cluster redshift, the difference in their observed B-R colour
is found to be 0.447\,mag (see Haines et al. \cite{hai03} for
details).  Thus we define as ``blue'' the galaxies with
(B-R)$_{\mathrm{corr}}$ $<$ -0.447.

The relation between D$_{n}$(4000) and (B-R)$_{\mathrm{corr}}$
(Fig.~\ref{BR_D4000}) shows that there is a clear separation between
blue and red galaxies and that we can adopt D$_{n}$(4000) = 1.45 as a
cut for the measurement of 4000 \AA \ \ break. This value corresponds
to that adopted by Balogh et al. (\cite{bal99}), for the
classification of galaxies in the CNOC sample.

   \begin{figure*}
   \centering
   \includegraphics[width=1.0\textwidth,bb= 3 400 577 696,clip]{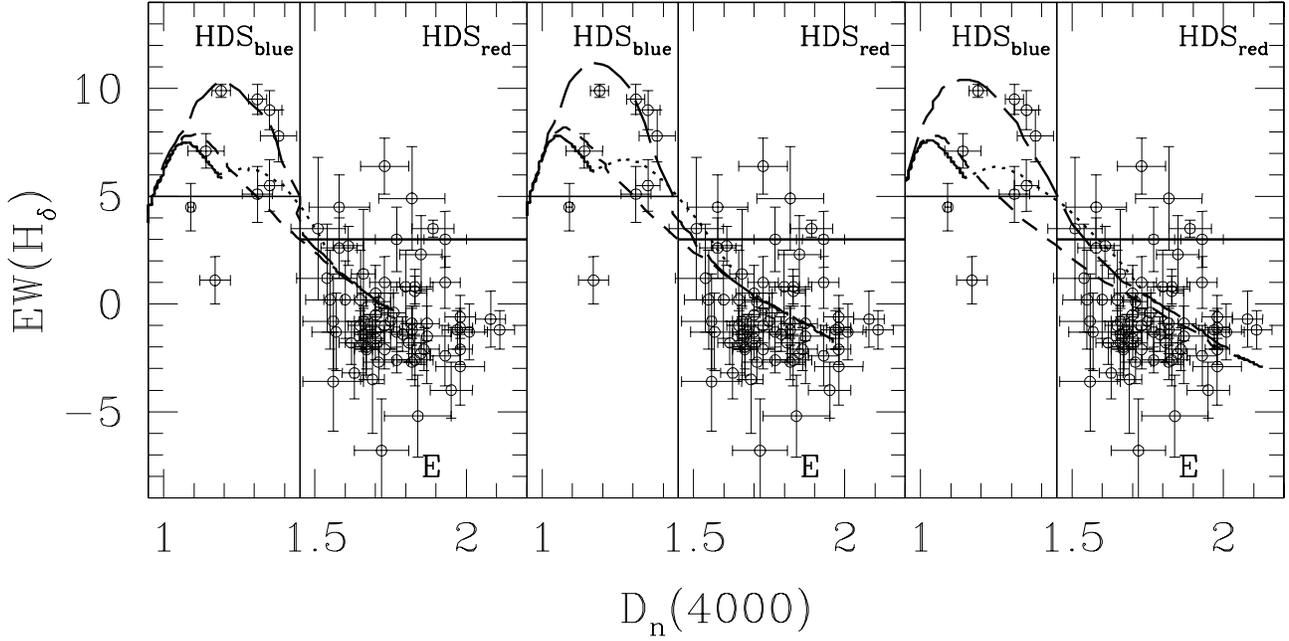}
    \caption{Equivalent width of H$_{\delta}$ in \AA \ \ versus D$_n$(4000)
      for 76 member galaxies compared to different GISSEL03
      models. Left panel shows models with Z=0.4Z$_\odot$, while the
      central and right panels refer to Z$_\odot$ and 2.5Z$_\odot$,
      respectively. In each panel, the short dashed curves represent a
      model with exponential declining SFR with $\tau$=1.0 Gyr, while
      for the continuous and dotted lines the decay parameter is
      $\mu$=0.01. For the last one, the star formation is truncated at
      t=10 Gyr. The long dashed lines represent a model with an initial
      100 Myr burst.}
   \label{hd_D4000}
   \end{figure*}

In Fig.~\ref{hd_D4000} we plot, in the D$_n$(4000)--EW(H$_\delta$)
plane, the data (open circles) of 76 galaxies, for which we have
information on both D$_n$(4000) and H$_\delta$, with superimposed the
output of four different models: i) a model with an initial burst of
star formation lasting 100 Myr (long dashed line); ii) a model with an
exponentially decaying SFR with decay parameter $\mu$=0.01 (continuous
line) lasting 11 Gyr and iii) truncated at 10 Gyr (dotted line); iv) a
model with an exponentially decaying SFR with time scale $\tau$ =1.0
Gyr (short dashed line).

For blue galaxies, to reach EW(H$_\delta$) $>$ 5.0 \AA, the
galaxy light must be dominated by A-- and early F--type stars, whereas
the presence of O-- and B--type stars, which have weak intrinsic
H$_\delta$ absorption, reduces this equivalent width. Moreover, the
equivalent width of H$_\delta$ is inversely related to the duration of
star formation. For this reason, the blue, H$_\delta$--strong galaxies
are expected to be the result of a short starburst or of a recently
terminated star formation (e.g. Dressler \& Gunn \cite{dre83}; Barger
et al. \cite{bar96}). By using models i) and iii) above, with
solar metallicity \footnote{\footnotesize Note that models with
Z=0.4Z$_\odot$ and Z=2.5Z$_\odot$ would lead to the same result (see
left and right panel in Fig.~\ref{hd_D4000}).}, when D$_n$(4000) is
equal to 1.45, the H$_\delta$ equivalent width is 5.0 \AA, so we
decided to use EW(H$_\delta$) $>$ 5.0 \AA \ \ as the selection
criterion for blue HDS.

On the contrary, normal early-type galaxies in the red half of the
plane generally show little or no signs of star formation, and can be
reproduced by a model with an exponentially decaying SFR with time
scale $\tau$ =1.0 Gyr. The model iv) with solar metallicity (see
the notes above) has EW(H$_\delta$) $<$ 3.0 \AA \ \ when
D$_n$(4000) $>$ 1.45. Therefore, for red HDS galaxies we adopted a
threshold of 3 \AA. Note that these selection criteria are equivalent
to those adopted by Balogh et al. (\cite{bal99}), derived by using
PEGASE models. None of the GISSEL03 models used here describes well
red HDS galaxies and require the introduction of other ingredients, as
discussed in Sect.~\ref{sec:6}.

5\% of cluster galaxies are classified as blue HDS galaxies and 7\% as
red HDS galaxies.

\subsection{Passive galaxies}

We define as passive, galaxies with no detectable emission lines
and EW(H$\delta$) $<$ 3.0 \AA. These criteria are chosen to select
galaxies without significant star formation, whose star formation
history can be modeled by the exponentially decaying rate with $\tau$
= 1.0 Gyr. Selected passive galaxies are mostly morphologically
ellipticals or S0s as is shown below.

81\% of member galaxies are classified as P galaxies.

\section{Comparison with structural properties}
\label{sec:5}

We analysed the spectral properties of disk and spheroidal galaxies,
separately (see La Barbera et al. \cite{lab03}). We define as
spheroids the objects with the Sersic index $n>2$, and the remaining
as disks. This corresponds to distinguishing between galaxies with a
low bulge fraction ($<$20\%) and those with a more prominent bulge
component (van Dokkum et al. \cite{van98}).

   \begin{figure*}
   \centering
   \includegraphics[width=0.7\textwidth,bb= 17 336 371 705,clip]{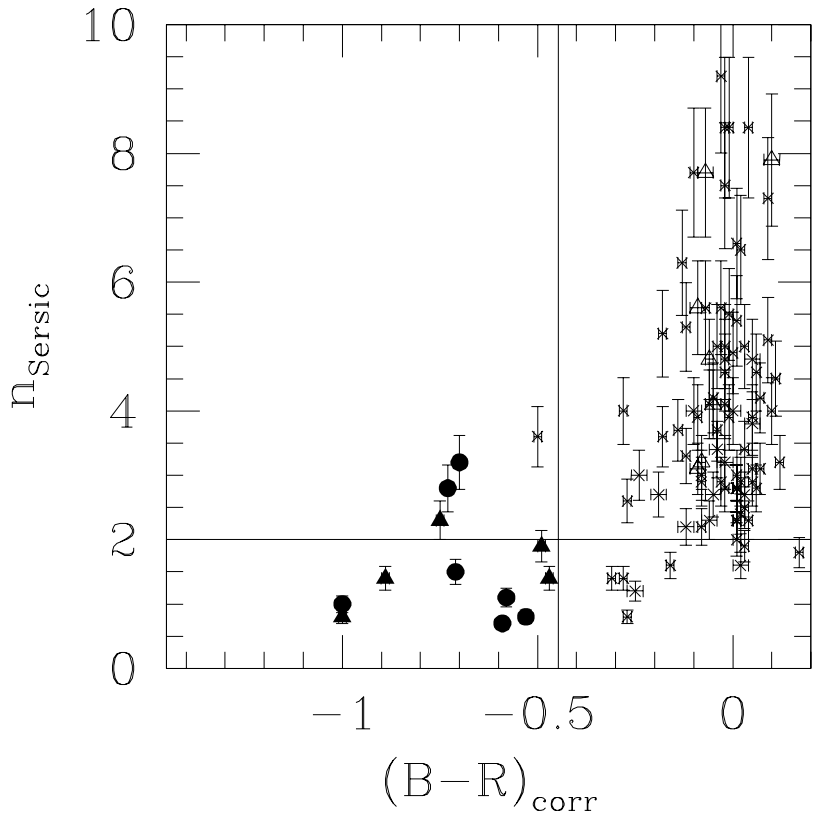}
	\vspace{-1.0cm}
    \caption{Comparison between B--R galaxy colours and Sersic indices
    (n$_{\mathrm{Sersic}}$). The vertical line marks the separation
    between blue and red galaxies (see text), while the horizontal line
    is the separation between disks and spheroids. Filled circles and
    triangles indicates ELG and HDS blue galaxies, respectively, while
    crosses and open triangles are for P and HDS red galaxies.}
   \label{BR_nsersic}
   \end{figure*}

\begin{figure*}
\centering
\includegraphics[width=0.8\textwidth]{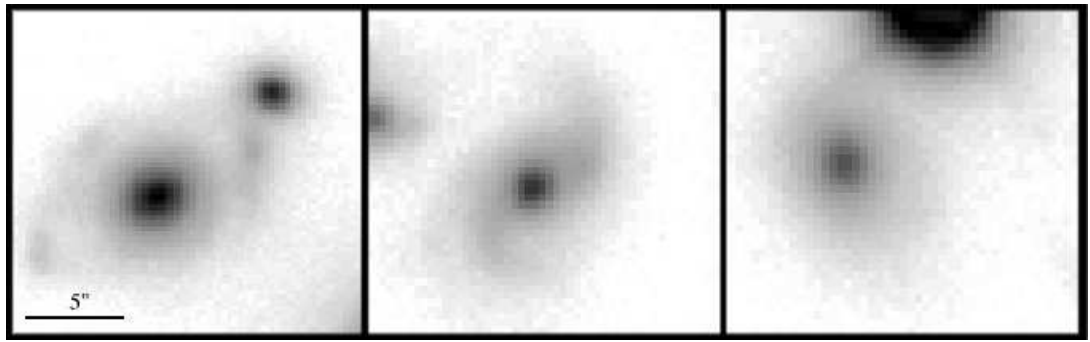}
\caption
{R--band image of the four galaxies discussed in the text, with North
at top and East to left. Scale is indicated in the left panel.}
\label{galaxies}
\end{figure*}

In Fig.~\ref{BR_nsersic} the Sersic index is plotted against the B-R
colour corrected for the effect of the CM relation, and different
symbols are used to distinguish the different galaxy spectral types.
This plot shows that there is a marked correspondence between spectral
and structural properties: blue disks are ELG or HDS$_{\mathrm{blue}}$
galaxies while red spheroids are P or HDS$_{\mathrm{red}}$ galaxies.
The converse is also true with some exceptions. Noticeably, one
HDS$_{\mathrm{blue}}$ and two ELG galaxies have $2<n<4$, and one P
galaxy has blue colour. By a visual inspection of R--band images, the
two ELGs show signs of spiral arms (Fig.~\ref{galaxies} left and
central panels). We note that one of those is also close to the P
galaxy with blue colour (see Fig.~\ref{BR_nsersic} and
Fig.~\ref{galaxies} left panel). Since the HDS$_{\mathrm{blue}}$
galaxy is close to a bright star (Fig.~\ref{galaxies} right panel),
its morphological classification could be wrong.

For the eight red P galaxies having Sersic index less than 2 we
suggest that they are a population of disk galaxies. van den Berg,
Pierce \& Tully (\cite{van90}), classifying 182 galaxies in the Virgo
cluster, have shown the existence of ``anemic'' spiral galaxies,
i.e. spiral galaxies deficient in neutral hydrogen. They have shown
that these observations may be accounted for by assuming that gas has
been stripped from the outer parts of these objects. Following van den
Berg \cite{van91} we recognize these galaxies as Ab-spirals and treat
them as a different sample of galaxies from that of ellipticals.

HDS$_{\mathrm{red}}$ galaxies have all Sersic indices greater than 3,
so they seem to be early-type galaxies.

\section{Distribution of ages}
\label{sec:6}

In order to follow the evolution of galaxies belonging to different
spectral classes, we compare the strengths of various spectral indices
in the observed galaxy spectra to stellar population models with
different star formation rates and metallicities.

Bruzual \& Charlot (\cite{bru03}) defined those indices that are most
suitable in their models to investigate galaxy populations. They
suggested two sets of spectral indices that should be fitted
simultaneously in order to break the age-metallicity degeneracy: one
of them being primary sensitive to the star formation history
(D$_n$(4000), H$_\beta$, H$_{\gamma A}$ and H$_{\delta A}$) and the
other primary sensitive to metallicity ([MgFe]$^\prime$,[Mg$_1$Fe] and
[Mg$_2$Fe]). The latter indices, in fact, are sensitive to metallicity
but do not depend sensitively on changes in alpha-element to iron
abundance ratios (see Thomas et al. \cite{tho03} and Bruzual \&
Charlot \cite{bru03} for details).

For these reasons we use models with different star formation
histories in order to fit simultaneously the observed strength of
D$_n$(4000), H$_{\delta A}$, H$_{\gamma A}$, H$_\beta$,
[MgFe]$^\prime$, [Mg$_1$Fe], and [Mg$_2$Fe].

   \begin{figure*}
   \centering
   \includegraphics[width=1.0\textwidth,bb= 3 400 577 696,clip]{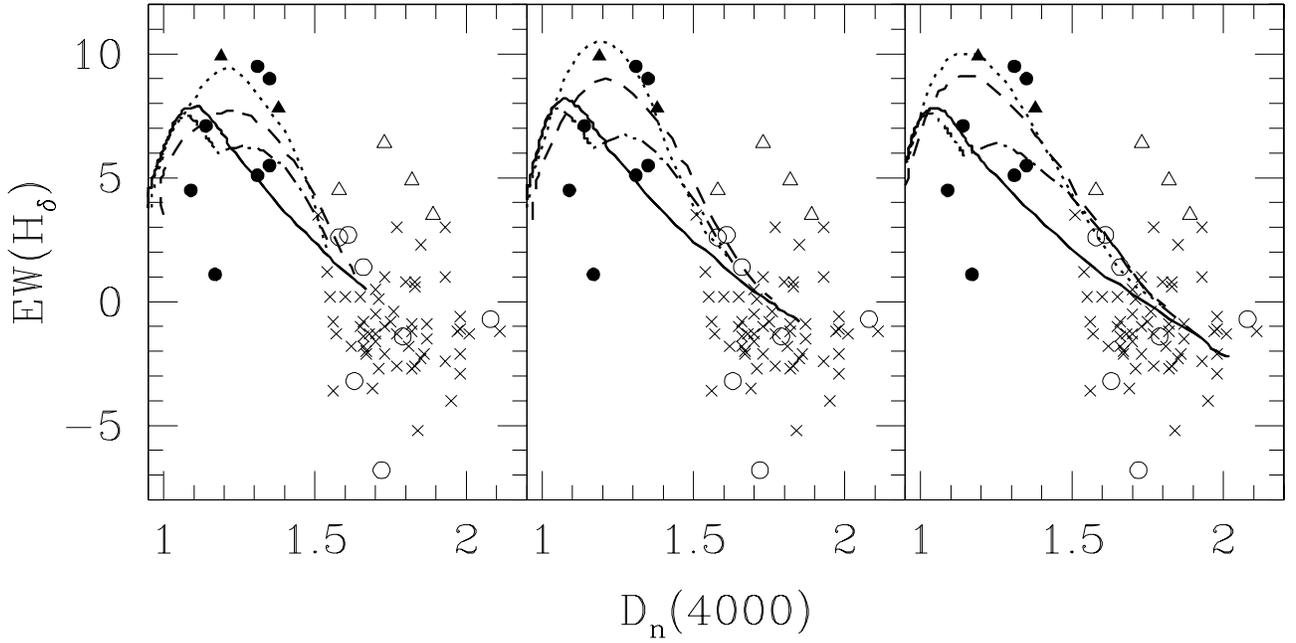}
    \caption{Equivalent width of H$_{\delta}$ in \AA \ \ versus the
      break at 4000 \AA \ \ for galaxies of different spectral types
      and GISSEL03 models. Filled and open circles, filled and open
      triangles, and crosses denote ELG, Ab--spiral,
      HDS$_{\mathrm{blue}}$, HDS$_{\mathrm{red}}$ and P galaxies,
      respectively. The left panel shows models with Z=0.4Z$_\odot$,
      central and right panels 1Z$_\odot$ and 2.5Z$_\odot$,
      respectively. In each panel it is plotted a model with
      exponential declining SFR with time scale $\tau$=1.0 Gyr lasting
      11 Gyr (continuous line), with superimposed a burst at t=9 Gyr,
      involving 10\% (dashed line) or 40\% (dotted line) of the total
      mass, and a model with continuous star formation truncated at
      t=10 Gyr (dot-dashed line). Note that models with
      continuous star formation stop at D$_n$(4000)= 1.2 as it is
      better shown by the continuous line in Fig.~\ref{hd_D4000}.}
   \label{hd_D4000_bis}
   \end{figure*}

We performed a chi-square minimization by using for each galaxy
the available spectral indices. Uncertainties on ages were estimated
performing Monte Carlo simulations. In each simulation, random shifts
were added to the best-fit EWs according to the estimated errors on
the spectral indices, and the uncertainty on age was obtained as the
standard deviation of the age distribution.

\begin{figure*}
   \centering
   \includegraphics[width=0.5\textwidth,bb= 7 428 304 710,clip]{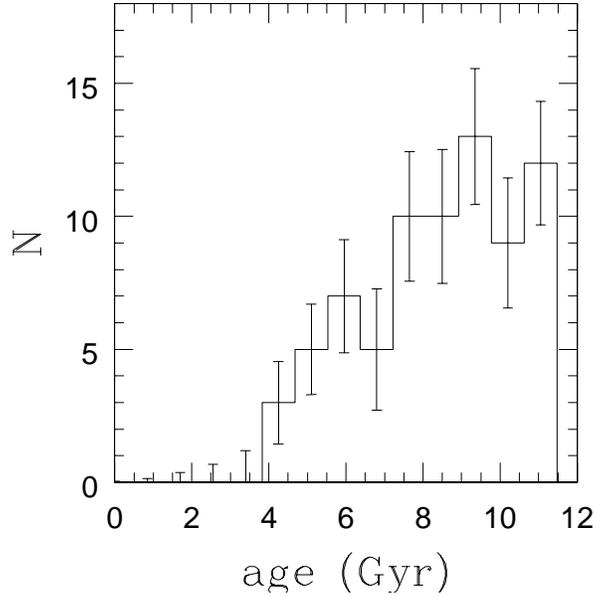}
    \caption{Distribution of best-fitted ages of P galaxies
      (continuous line), by using a model with exponential declining
      SFR with time scale $\tau$=1.0 Gyr. Errors on ages are obtained
      performing 1000 simulations accounting for
      equivalent width measurement errors (see text)}.
   \label{eta_tau1}
   \end{figure*}

P-galaxies (crosses in Fig.~\ref{hd_D4000_bis}) exhibit
moderate dispersion in D$_n$(4000), caused by a combination of age and
metallicity effects and little star formation. They are well described
by an exponentially declining star formation with time scale
$\tau$=1.0 Gyr (Merluzzi et al. \cite{merl03} and references
therein). Assuming this SFR and metallicities Z=0.4
Z$_\odot$,Z$_\odot$,2.5 Z$_\odot$, the distribution of ages for
early-type galaxies, (Fig.~\ref{eta_tau1}), shows that the age of the
bulk of ellipticals in ABCG\,209 is greater than $\sim$ 7 Gyr and that
there is a tail toward younger ages up to $\sim$ 4 Gry. This
translates into a formation epoch z$_f \gtrsim$ 1.6 for the majority
of ellipticals and z$_f \gtrsim$ 0.7 for the younger population.  The
KMM algorithm (cf. Ashman et al. \cite{ash94} and refs. therein)
suggests that a mixture of two Gaussians (with $\mathrm{n_1=19}$, and
$\mathrm{n_2=55}$ members) is a better description of this age
distribution (although only at $\sim 90\%$ c.l.).  In this case, the
mean ages of the two distributions are $\mathrm{t_1 = 5.8 \pm 0.8}$
Gyr, and $\mathrm{t_{2} = 9.2 \pm 1.2}$ Gyr respectively. Thus, 74.3\%
of early-type galaxies seem to be coeval and to be formed early during
the initial collapse of the cluster, at z$_f \gtrsim$ 3.5. The
remaining 25.7\% is constituted by a younger galaxy population, that
could be formed later (z$_f \gtrsim$ 1.2) or could have experienced an
enhancement in the star formation rate $\sim$ 8.5 Gyr ago. This is in
remarkable agreement with the trend with redshift of the total stellar
mass in the cluster galaxies, derived from the colour-magnitude
relation by Merluzzi et al. \cite{merl03} (see their Fig.~10).

As shown in Fig.~\ref{hd_D4000_bis} (filled circles), ELG galaxies can
be fitted either by a model with truncated star formation history and
by a model with a burst involving the 40\% of the total mass
(Fig.~\ref{hd_D4000_bis}), whereas the two HDS$_{\mathrm{blue}}$
galaxies, for which we have the measurement of both D$_n$(4000) and
H$_\delta$ equivalent widths, seem to be reproduced only by a
model with a short starburst. The starburst is assumed to begin at t=9
Gyr lasting for 0.1 Gyr. An equivalent width of H$_\delta$ greater
than 5 \AA \ \ implies that a starburst occurred in the galaxy before
star formation was quenched, otherwise quiescent star formation
activity would produce a spectrum with weaker Balmer lines (e.g.,
Couch \& Sharples \cite{cou87}; Poggianti \& Barbaro \cite{pog96}), as
showed also by the comparison of the two models (dotted and dot-
dashed lines) in Fig.~\ref{hd_D4000_bis}. HDS$_{\mathrm{blue}}$
galaxies have EWs stronger than 5 \AA, and thus could be
post-starburst galaxies (but see below). The strength of the lines
and the colour of these galaxies indicate an interruption of the star
formation within the last 0.5 Gyr, with a strong starburst preceding
the quenching of star formation (see also Poggianti et
al. \cite{pog99}; Poggianti \& Barbaro \cite{pog96}).

In contrast, the typical time elapsed since the last star formation in
the HDS$_{\mathrm{red}}$ will be in the range 1-2 Gyr, since this
population exhibits weaker H$_{\delta}$ equivalent widths. However,
the observed H$_{\delta}$ values for HDS$_{\mathrm{red}}$ tend to be
larger than those predicted by both starburst and truncated star
formation models, except from one galaxy that is well reproduced by a
burst model (observed $\sim$ 1 Gyr after the burst). This problem was
pointed out by several authors (e.g. Couch \& Sharples \cite{cou87},
Morris et al. \cite{mor98}, Poggianti \& Barbaro \cite{pog96}, Balogh
et al. \cite{bal99}) and will persist if photometric colours are
considered instead of D$_n$(4000). Models with IMFs biased toward
massive stars are able to reproduce better these data. However, the
stellar population resulting from such a burst is very short lived
($\sim$ 300 Myr), thus it is unlikely that many of these red galaxies
are undergoing such bursts.  Stronger H$_{\delta}$ equivalent
widths may also be produced by temporal variations in the internal
reddening in truncated or burst models. Balogh et al. (\cite{bal99})
have shown that by reddening the model it is possible to recover the
observed data. This supports the idea that dust obscuration may play
an important role in the appearance of these spectra. Recently, in
fact, Shioya, Bekki \& Couch (\cite{shi04}) demonstrates that the
reddest HDS galaxies can only be explained by truncated or starburst
models with very heavy dust extinction ($A_V > $ 0.5 mag).

\begin{figure*}
   \centering
   \includegraphics[width=1.0\textwidth,bb= 3 400 577 696,clip]{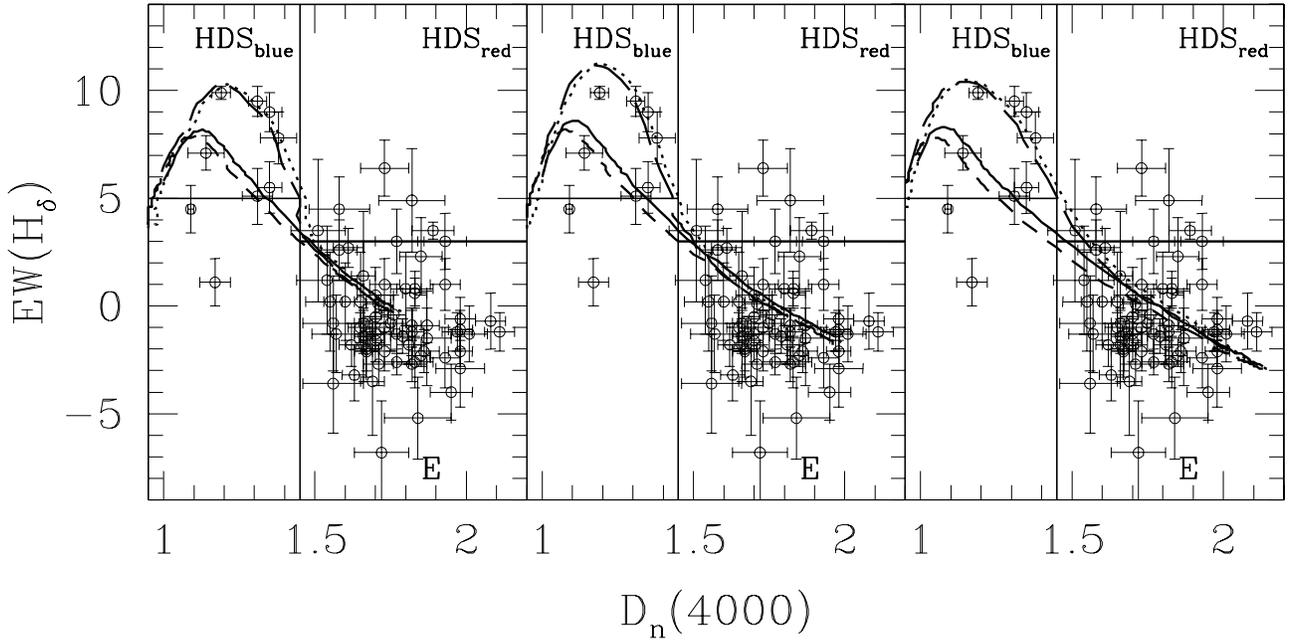}
    \caption{Equivalent width of H$_{\delta}$ in \AA \ \ versus
      D$_n$(4000) for 76 member galaxies compared to different
      GISSEL03 models. Left panel shows models with Z=0.4Z$_\odot$,
      while the central and right panels refer to Z$_\odot$ and
      2.5Z$_\odot$, respectively. In each panel, the short and the
      long dashed curves represent a model with exponential declining
      SFR with $\tau$=1.0 Gyr and with an initial 100 Myr burst
      without dust respectively. The continuous and the dotted lines
      draw the same models with dust extinction with
      $\tau^{\mathrm{BC}}_{\mathrm{V}}$ = 1.0 and
      $\tau^{\mathrm{ISM}}_{\mathrm{V}}$ = 0.5.}
   \label{dust}
   \end{figure*}

We run Bruzual \& Charlot (\cite{bru03}) models with dust
extinction, in order to check if the typical amount of dust content of
galaxies can explain the red colour of HDS$_{\mathrm{red}}$
galaxies. We used the standard values of dust extinction:
$\tau^{\mathrm{BC}}_{\mathrm{V}}$ = 1.0 and
$\tau^{\mathrm{ISM}}_{\mathrm{V}}$ = 0.5 reported by Charlot \& Fall
(\cite{cha00}). As shown in Fig.~\ref{dust}, the typical amount of
dust content in galaxies is inadequate to explain the colour of
HDS$_{\mathrm{red}}$ galaxies.

\section{Dynamical scenario}
\label{sec:7}

The analysis of the velocity distribution for all member galaxies of
ABCG\,209 (Mercurio et al. \cite{mer03a}) shows the existence of a
peak at $<z>=0.2090\pm 0.0004$ with a line--of--sight velocity
dispersion $\sigma_v=1394^{+88}_{-99}$ \kss.

In order to check for possible variations of $<z>$ and $\sigma_v$ for
different spectral types we analysed separately ELG,
HDS$_{\mathrm{blue}}$, Ab-spirals, HDS$_{\mathrm{red}}$, and P
galaxies. All galaxy populations show consistent mean redshift
suggesting that they are really cluster populations and not merely
foreground or background objects. On the other hand, the velocity
dispersions of different classes show significant differences (see
Table~\ref{vel_disp}).

\begin{table}

     \caption[]{Redshifts and velocity dispersion of different
     spectral classes. The biweight estimators (Beers et
     al. \cite{bee90}) were used to derive the values}.
    \label{vel_disp}
    $$
           \begin{array}{c c c c }
            \hline
            \noalign{\smallskip}
    \mathrm{Class} & \mathrm{N_{gal}} &  \mathrm{Redshift} & \mathrm{\sigma_{v}} \\
            \noalign{\smallskip}
            \hline
            \noalign{\smallskip}
\mathrm{ALL}        & 112 & 0.2090\pm0.0004 & 1394_{-99}^{+88}\\
\mathrm{P}          &  74 & 0.2086\pm0.0005 & 1323_{-94}^{+127}\\
\mathrm{HDS_{red}}  &   7 & 0.2087\pm0.0005 & 339_{-46}^{+91}\\
\mathrm{Ab-spirals}  &   8 & 0.2107\pm0.0017 & 1295_{-102}^{+340}\\
\mathrm{HDS_{blue}} &   5 & 0.2071\pm0.0044 & 2257_{-65}^{+1085}\\
\mathrm{ELG}        &   7 & 0.2070\pm0.0038 & 2634_{-53}^{+712}\\
            \noalign{\smallskip}
            \hline
         \end{array}
     $$
   \end{table}

P galaxies define the mean redshift and the velocity dispersion of the
cluster. The very low value of the velocity dispersion of
HDS$_{\mathrm{red}}$ galaxies indicates that they could constitute a
group of galaxies located at the cluster redshift. These
post-starburst galaxies could be the remnant of the core of an
infalling clump, that have suffered significant ram--pressure
stripping in crossing the cluster core, preceded by an instantaneous
burst of star formation. 

The groups of ELG and HDS$_{\mathrm{blue}}$ galaxies exhibit very
high velocity dispersions.  The high velocity dispersions of these
populations of disk-dominated galaxies indicate that they do not
constitute a single infalling bound group but that they are the result
of the growing of the cluster through the accretion of field galaxies.
They have very similar spectral, photometric and structural properties
and only slight difference in the velocity dispersion. On the other
hand, they lie in regions with different density (see below).

Figure~\ref{spat_distr} shows the spatial distribution of different
spectral classes. Taking into account the SE-NW elongation of the
cluster, ELGs (filled circles) seem to be uniformly distributed in the
outer parts of the cluster, while blue HDS galaxies (filled triangles)
show a preferential location within the cluster. They are concentrated
around the centre of the cluster along a direction perpendicular to
the cluster elongation. P galaxies (crosses) are uniformly
distributed over the whole cluster plane and red HDS galaxies (open
triangles) lie around the centre along the cluster
elongation. Ab--spirals (open circles) avoid the highest density
regions, in agreenment with those found by Goto et
al. (\cite{got03c}).

   \begin{figure*}
   \centering
   \includegraphics[width=0.7\textwidth]{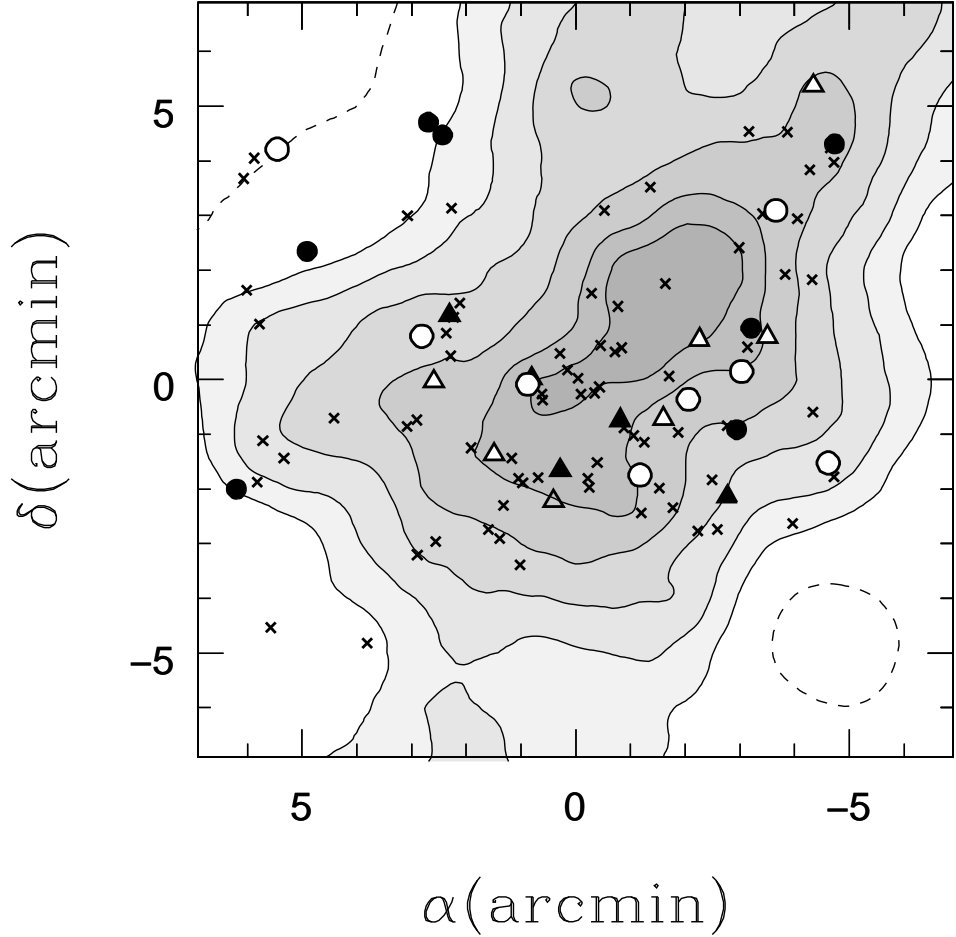}
    \caption{Spatial distribution of different spectral
     classes. Filled and open circles, filled and open triangles, and
     crosses denote ELG, Ab--spiral, HDS$_{\mathrm{blue}}$,
     HDS$_{\mathrm{red}}$ and P galaxies, respectively. The plot is
     centred on the cluster center. The dashed contour corresponds to
     the field galaxy number density, while the solid contours
     correspond respectively to the background subtracted 2, 3.3, 5,
     7.5, 10, and 12.5 galaxies arcmin$^{-2}$}
   \label{spat_distr}
   \end{figure*}

By using the information about the phase-space distributions, it is
possible to find segregations of the various galaxy populations.
Biviano et al. (\cite{biv02}) searched for segregation by comparing
(R,v)-distributions in the clusters of the ESO Nearby Abell Cluster
Survey, using the combined evidence from projected positions and
relative velocities (they may not be independent). They also stressed
the importance of the estimate of the projected clustercentric
distance R, in order to make the comparison as unbiased as possible.

The spatial distribution of galaxies in ABCG\,209 shows a complex
structure, characterised by an elongation in the SE-NW direction (see
Mercurio et al. \cite{mer03a}). To measure the effect of the cluster
environment on the galaxy population, their spectral properties should
be measured against the parameter most related to the cluster
structure, which is the local galaxy number surface density rather
than cluster-centric radius.

The local surface density (LD) of galaxies was determined across the
CFHT images by Haines et al. (\cite{hai03}). We performed the
comparison of (LD,v)-distributions by means the two--dimensional
Kolmogorov--Smirnov test (hereafter 2DKS--test; Press et
al. \cite{pre92}). The 2DKS test is relatively conservative: if it
indicates that two distributions have a high probability of not being
drawn from the same parent population, other tests (e.g. Rank-Sum
tests or Sign Tests) indicate the same thing, while the contrary is
not always true. Moreover, the 2DKS test turns out to be reliable with
relatively small samples (e.g. Biviano et al. \cite{biv02}).

According to the 2DKS--test, we find significant evidence for
segregation of ELGs with respect to both Ps (99.5\%) and
HDS$_{\mathrm{blue}}$ galaxies (97.6\%), where ELGs lie in regions
with lower density than P and HDS$_{\mathrm{blue}}$ galaxies.
Moreover the distributions of HDS$_{\mathrm{blue}}$ and
HDS$_{\mathrm{red}}$ galaxies differ at 90.7\% c.l. and
HDS$_{\mathrm{blue}}$ and Ab-spirals at 92.0\%. These results show the
effects of the cluster environment on the spectral properties of their
constituents and support the scenario in which spirals are
accreted into the cluster from the field, then star formation is
stopped, the galaxies become gas-deficient, and eventually undergo a
morphological transformation (Couch et al. \cite{cou98}). This
scenario may also account for the existence of anemic spirals
(Ab-spirals in Table~\ref{vel_disp}), whose velocity dispersion is
fully consistent with those of the cluster.

\section{Summary and Discussion}
\label{sec:8}

In order to investigate the nature of the galaxy population in
ABCG\,209, we have performed a detailed study of the photometric and
spectroscopic properties of luminous member galaxies (R $\lesssim$
20.0).  The primary goal of this analysis was to study the relation
between cluster dynamics and evolution of stellar populations, and the
effects of cluster environment on the properties of galaxies.

ABCG\,209 is undergoing strong dynamical evolution with the merging of
at least two clumps along the SE-NW direction, as indicated by i) the
presence of a velocity gradient along a SE-NW axis, ii) the elongation
of X-ray contour levels in Chandra images, and iii) the presence of
substructures manifest both in the detection of two peaks separated by
50 arcsec in the X-ray flux and in the analysis of the velocity
distribution, that showed three sub-clumps (Mercurio et
al. \cite{mer03a}). Moreover, the young dynamical state of the cluster
is indicated by the possible presence of a radio halo (Giovannini,
Tordi \& Feretti \cite{gio99}), which has been suggested to be the
result of a recent cluster merger, through the acceleration of
relativistic particles by the merger shocks (Feretti \cite{fer02}).

The analysis of the spectroscopic and photometric properties of 102
cluster members has shown the presence of five different types of
galaxies: i) passive evolving galaxies (P), which exhibit red colours
and no emission lines, ii) emission line galaxies (ELG), which are
blue and have prominent emission lines, iii) strong blue H$_\delta$
galaxies (HDS$_{\mathrm{blue}}$), that are characterized by the
presence of strong H$_\delta$ absorption lines ( EW(H$_\delta$) $>$ 5
\AA) and blue colours , iv) strong red H$_\delta$ galaxies
(HDS$_{\mathrm{red}}$) galaxies, having EW(H$_\delta$) $>$ 3 \AA \ \
and red colours; v) anemic spirals (Ab-spirals), that have the same
spectral properties of passive evolving galaxies, but are
disk-dominated systems.

P galaxies represent $\sim$ 74\% of the cluster population, and lie
mainly in high density regions. Their age distribution is
characterized by the presence of an ``old population'' formed early
during the initial collapse of the cluster at ages greater than 7 Gyr,
and another one, younger, that could be formed later or could have
experienced an enhancement in the star formation rate.

ELGs lie in low density regions and have high line--of--sight velocity
dispersion. Both the spatial position and the velocity dispersion
suggest that this is a population of galaxies that has recently fallen
into the cluster. This infall has induced a truncation of the star
formation with, possibly, an associated short starburst. In fact, as
shown in Fig.~\ref{hd_D4000_bis} (compare also with
Fig.~\ref{hd_D4000}), only one ELG galaxy is consistent with
continuous the star formation model before truncation.

These results are understandable in terms of cosmological models of
structure formation, in which galaxies form earliest in the
highest-density regions corresponding to the cores of rich
clusters. Not only do the galaxies form earliest here, but the bulk of
their star formation is complete by $z\sim1$, at which point the
cluster core is filled by shock-heated virialised gas which does not
easily cool or collapse, suppressing the further formation of stars
and galaxies (Blanton et al. \cite{blanton99};
\cite{blanton00}). Diaferio et al. (\cite{dia01}) shows that as mixing
of the galaxy population is incomplete during cluster assembly, the
positions of galaxies within the cluster are correlated with the epoch
at which they were accreted. Hence galaxies in the cluster periphery
are accreted later, and so have their star formation suppressed later,
resulting in younger mean stellar populations. It should be considered
however that this is a small-scale affect, and that this result in
fact confirms that the red sequence galaxy population is remarkably
homogeneous across all environments.

A similar scenario is suggested also by the velocity dispersion and
the spectral properties of HDS$_{\mathrm{blue}}$ galaxies, which,
however, are found to be aligned in a direction perpendicular to the
cluster elongation, that coincides with the elongation of the X-ray
contour levels in the Chandra images (see Fig.~14 in Mercurio et
al. \cite{mer03a}). As already pointed out by Poggianti et
al. (\cite{pog04}) analysing the position of the strongest k+a
galaxies (having EW(H$_\delta$) $>$ 5 \AA) in Coma, the correlation
between the location of the post-starburst galaxies and the
substructures in the intracluster medium strongly suggest that the
truncation of the star formation activity in these galaxies, and
possibly the previous starburst, could be the result of an interaction
with the hot intracluster medium. Thus, the origin of
HDS$_{\mathrm{blue}}$ could be a cluster related and, in particular,
an ICM-related phenomenon, closely connected with the dynamical state
of the cluster (see Poggianti et al. \cite{pog04}). On the other hand,
other possible explanations are reported in literature for the origin
of HDS galaxies (e.g. Zabludoff et al. \cite{zab96}, Balogh et
al. \cite{bal99}, Goto \cite{got03d}, Quintero et
al. \cite{qui03}). In particular, an open question is whether the
burst is related to galaxy--galaxy or galaxy--cluster interactions.

HDS$_{\mathrm{red}}$ galaxies are distributed along the elongation of
the cluster, mainly in intermediate density regions and have a low
velocity dispersion. According to the evolution models, the presence
of strong H$_\delta$ absorption line in their spectra indicates that
these galaxies have experienced a short starburst of star formation in
the past few Gyr. The measured D$_{\mathrm{n}}$(4000) of these
galaxies can be made consistent with short starburst models only in
presence of a substantial reddening by dust. In the starburst model,
EW(H$_{\delta}$) and D$_n$(4000) decline on a timescale of $\sim$ 2
Gyr after the burst has ceased, regardless of their SFR before the
burst. This implies that, in the galaxies we observe, the burst has
occurred no more than 4.5 Gyr ago.

As also indicated by the low value of the velocity dispersion,
HDS$_{\mathrm{red}}$ galaxies could be the remnant of the core of an
infalling clump of galaxies, that have experienced a merger with the
main cluster. This merger may have induced a time--dependent tidal
gravitational field that stimulated non--axisymmetric perturbations in
the galaxies, driving effective transfer of gas to the central
galactic region and, finally, triggering a secondary starburst in the
central part of these galaxies (Bekki \cite{bek99}). The second peak
of the X-ray distribution could be related to the presence of this
small galaxy group.

Considering that the mass is proportional to the cube of the measured
velocity dispersion (see eq. (4) and (11) of Girardi et
al. \cite{gir98}), the ratio between the mass of the cluster as
constituted only by P galaxies and the group of HDS$_{\mathrm{red}}$
galaxies is $\sim$ 0.017. Taffoni et al. (\cite{taf03}) studied the
evolution of dark matter satellites orbiting inside more massive
haloes, by using semi-analytical tools coupled with high-resolution
N-body simulations. They explored the interplay between dynamic
friction and tidal mass loss/evaporation in determining the final fate
of the satellite, showing that, for satellites of intermediate mass
(0.01 M$_\mathrm{h} <$ M$_\mathrm{s,0} < $ 0.1 M$_\mathrm{h}$, where
M$_\mathrm{s,0}$ is the initial mass of the satellite and
M$_\mathrm{h}$ is the mass of the main halo), the dynamical friction is
strong and drives the satellite toward the centre of the main halo,
with significant mass loss. The final fate depends on the
concentration of the satellite, relative to that of the main
halo. Low--concentration satellites are disrupted, while
high--concentration satellites survive, with a final mass that
depends on the decay time.

Although both HDS$_{\mathrm{blue}}$ and HDS$_{\mathrm{red}}$ are
characterized by the presence of strong Balmer lines, their colours
and structural parameters show that they are two different galaxy
populations. Deriving the Sersic index, we found that
HDS$_{\mathrm{blue}}$ galaxies are disk-dominated galaxies while
HDS$_{\mathrm{red}}$ are spheroids. On the contrary, Poggianti et
al. (\cite{pog04}) found that, with the exception of two blue k+a
galaxies compatible with a de Vaucouleurs' profile, all the other blue
and red k+a galaxies in the Coma cluster have exponential or steeper
profiles. This apparent contradiction may be overcome by considering
the time for a HDS$_{\mathrm{blue}}$ galaxy to become red due to the
damping of star formation is shorter than the difference in look-back
time between ABCG\,209 and Coma. It is thus conceivable that the
HDS$_{\mathrm{blue}}$ galaxies we observe at z$\sim$0.21 will turn
into HDS$_{\mathrm{red}}$ galaxies observed Coma.

All the presented results support an evolutionary scenario in which
ABCG\,209 is characterized by the presence of two components: an old
galaxy population, formed very early (z$_f \gtrsim $ 3.5) and a
younger population of infalling galaxies. Moreover, this cluster may
have experienced, 3.5-4.5 Gyr ago, a merging with an infalling
galaxy group.  The merger of the cluster with the infalling group
could have also powered the observed radio halo. In fact, merger
activity and high ICM temperature may be responsible for producing a
radio halo (Liang et al. \cite{lia00}), because merging can provide
enough energy to accelerate the electrons to relativistic energies,
give rising to non-thermal emission. After the shock disappeared,
radio halos may be maintained in situ by electron acceleration in the
residual turbulence.

\section*{Appendix A}
\label{sec:A}

The equivalent width is defined by:

\begin{equation}
\mathrm{EW} = \sum^{N}_{i=1} \frac{\mathrm{F_{Ci}} - \mathrm{F_i}}
{\mathrm{F_{Ci}}} \Delta \lambda \ , 
\label{eq1}
\end{equation}

where F$_i$ is the flux in the pixel $i$, $N$ is the number of pixels
in the integration range, $\Delta \lambda$ is the dispersion in
\AA/pixel and F$_{Ci}$ is the straight line fitted to the blue and the
red pseudo--continuum intervals, evaluated in the pixel $i$ (pixels
with values more than 3$\sigma$ away from the continuum level were
rejected).

An index measured in magnitudes is:

\begin{equation}
\mathrm{Mag} = - 2.5 \mathrm{log} \left[1 - \frac{\mathrm{EW}}
{\mathrm{(\lambda_2-\lambda_1)}}\right] \ , 
\label{eq2}
\end{equation}

where $\lambda_1$ and $\lambda_2$ are the wavelength limits of the
feature bandpass.

The errors of the equivalent width measurement are estimated from
the following relation (Czoske et al. \cite{czo01}):

\begin{equation}
\sigma^2_{EW}=\left(\frac{\mathrm{S}}{\mathrm{N}}\right)^{-2} \left[(N \Delta \lambda - EW) \Delta \lambda + \frac{(N \Delta \lambda - EW)^2}{N_1+N_2} \right] \ ,
\label{eq4}
\end{equation}
where S/N is the signal--to--noise ratio evaluated in the feature bandpass.

The local S/N for each galaxy is obtained by dividing the galaxy
spectrum by its associated noise spectrum, as determined by adding in
quadrature the Poisson noise (obtained through the IRAF task APALL)
and the read-out noise of EMMI. The result is then fitted to remove
residual features, and averaged over the wavelength range of the
spectral feature to obtain the S/N ratio.
 
The catalogue of spectroscopic measurements is presented in Table
~\ref{catalogue}. 

\begin{landscape}
\begin{table}

        \caption[]{\footnotesize Spectroscopic data. Running number
        for galaxies in the presented sample (see also Table
        \ref{phot_cat}), ID (Col.~1); strength of the 4000 \AA \ \
        break (Col.~2); Cols.~3-15: equivalent widths. A value of
        '0.0' for the equivalent width for an emission-line indicates
        that no emission is observed, whereas '....'  indicates that
        the equivalent width could not be measured as the line does
        not lie within the available wavelength range. The spectral
        classification adopted in the present paper is listed in the
        last column. ELG: emission line galaxies;
        HDS$_{\mathrm{blue}}$: H$_\delta$ strong galaxies with blue
        colour; HDS$_{\mathrm{red}}$: H$_\delta$ strong galaxies with
        red colour; P: passive evolving galaxies; Ab-spirals: passive
        evolving disk-dominated galaxies (n$_{\mathrm{Sersic}} \le$
        2). Galaxy 62 presents very strong [OII], [OIII] and
        H$_\alpha$ emission, that could suggest the presence of an
        AGN. However, our data do not allow a realiable
        classification, leading us to exclude this object from the
        analysis.}

         \label{catalogue}
{\footnotesize
              $$ 
           \begin{array}{c c c c c c c c c c c c c c c c }
            \hline
            \noalign{\smallskip}
            \hline
            \noalign{\smallskip}
\mathrm{ID} & 
\mathrm{D_n(4000)} & \mathrm{[OII]} & \mathrm{H_{\delta A}} & 
\mathrm{H_{\gamma A}} & \mathrm{Fe4531} & \mathrm{H_{\beta}} & \mathrm{[OIII]}
& \mathrm{Fe5015} & \mathrm{Mg_1} & \mathrm{Mg_2} & \mathrm{Mg}b & 
\mathrm{Fe5270} & \mathrm{Fe5335} & \mathrm{H_{\alpha}} & \mathrm{Spectral}\\
 & \AA & \AA & \AA & \AA & \AA & \AA & \AA & \AA & 
\mathrm{mag} & \mathrm{mag} & \AA & \AA & \AA & \AA & \mathrm{class}\\
            \hline
            \noalign{\smallskip}   
  1&1.14\pm0.06&-10.1\pm 1.2&  7.1\pm 0.8&  0.3\pm 0.8&  4.7\pm 1.0& -0.5\pm 0.6& -0.3\pm 0.4&  5.2\pm 1.2&0.07\pm0.01&0.21\pm0.02&  3.4\pm 0.5&  2.7\pm 0.6&  2.0\pm 0.7& -10.8\pm  0.9& \mathrm{ELG}\\       
  2&1.73\pm0.03&  0.0       & -1.0\pm 0.8& -4.2\pm 0.7&  4.3\pm 0.7&  1.6\pm 0.5&  0.0       &  6.1\pm 0.9&0.16\pm0.01&0.30\pm0.01&  3.6\pm 0.4&  3.6\pm 0.5&  2.7\pm 0.6&   2.5\pm  0.4& \mathrm{P} \\        
  3&1.93\pm0.05&  0.0       &  1.0\pm 1.2& -2.8\pm 1.1&  2.9\pm 1.1&  2.5\pm 0.7&  0.0       &  7.4\pm 1.3&0.15\pm0.02&0.33\pm0.02&  5.6\pm 0.6&  5.4\pm 0.7&  2.2\pm 0.9&   2.7\pm  0.6& \mathrm{P} \\        
  4&  ....     &  ....      & -4.0\pm 2.2&  1.4\pm 1.8&  2.4\pm 2.1&  4.2\pm 1.3&  0.0       & 11.3\pm 2.5&0.12\pm0.03&0.18\pm0.04&  3.9\pm 1.3&  5.5\pm 1.4&  3.4\pm 1.6&   5.5\pm  1.2& \mathrm{P} \\        
  5&1.72\pm0.09&  ....      & -6.8\pm 2.4& -1.8\pm 1.6& -0.4\pm 1.8&  5.3\pm 1.0&  0.0       &  5.4\pm 2.1&0.10\pm0.02&0.27\pm0.03&  1.3\pm 1.0&  5.9\pm 1.0&  2.0\pm 1.3&  -4.2\pm  1.1& \mathrm{Ab-spiral} \\        
  6&1.89\pm0.07& 0.0        &  3.5\pm 0.4& -9.0\pm 0.5&  2.6\pm 0.5&  2.8\pm 0.3&  0.0       & 10.6\pm 0.7&0.19\pm0.01&0.44\pm0.01&  6.9\pm 0.3&  6.4\pm 0.4&  5.8\pm 0.5&   3.6\pm  0.4& \mathrm{HDS_{red}} \\        
  7&1.70\pm0.03& 0.0        & -1.3\pm 0.7& -3.0\pm 0.6&  4.1\pm 0.7&  1.9\pm 0.4&  0.0       &  5.1\pm 0.8&0.16\pm0.01&0.34\pm0.01&  4.3\pm 0.4&  3.1\pm 0.4&  2.2\pm 0.5&   0.3\pm  0.4& \mathrm{P} \\        
  8&1.73\pm0.05&  0.0       &  1.0\pm 1.2& -3.9\pm 1.0&  4.7\pm 1.0&  1.3\pm 0.6&  0.0       &  7.5\pm 1.2&0.19\pm0.02&0.36\pm0.02&  5.2\pm 0.6&  3.9\pm 0.7&  2.3\pm 0.8&   0.4\pm  0.6& \mathrm{P} \\        
  9&  ....     &  ....      & -1.6\pm 1.6& -0.3\pm 1.3&  4.8\pm 1.4&  1.1\pm 0.9&  0.0       &  6.0\pm 1.7&0.17\pm0.02&0.34\pm0.03&  4.1\pm 0.8&  4.4\pm 0.9&  6.2\pm 1.0&   0.4\pm  0.8& \mathrm{P} \\        
 10&1.66\pm0.06&  ....      & -1.8\pm 1.7& -3.5\pm 1.4&  4.2\pm 1.4&  2.0\pm 1.0&  0.0       &  7.0\pm 1.9&0.15\pm0.02&0.31\pm0.03&  4.9\pm 0.8&  3.1\pm 0.9&  4.3\pm 1.1&   4.6\pm  0.7& \mathrm{P} \\        
 11&1.77\pm0.04&  0.0       & -1.3\pm 0.9& -4.6\pm 0.8&  4.4\pm 0.9&  2.8\pm 0.5&  0.0       &  5.4\pm 1.1&0.17\pm0.01&0.32\pm0.02&  5.1\pm 0.5&  3.1\pm 0.6&  1.9\pm 0.6&   1.8\pm  0.5& \mathrm{P} \\        
 12&  ....     &  ....      & -2.7\pm 1.7& -8.4\pm 1.5&  4.4\pm 1.6&  4.0\pm 1.0&  0.0       &  7.5\pm 2.0&0.24\pm0.03&0.31\pm0.03&  4.0\pm 0.9&  8.1\pm 1.0&  2.8\pm 1.2&   1.7\pm  1.0& \mathrm{P} \\        
 13&  ....     &  ....      &  2.8\pm 1.8& -2.8\pm 1.6&  4.4\pm 1.7&  1.0\pm 1.1&  0.0       &  8.3\pm 2.0&0.14\pm0.03&0.27\pm0.03&  2.8\pm 1.0&  7.1\pm 1.0&  4.7\pm 1.2&   1.4\pm  0.9& \mathrm{P} \\ 	  
 14&2.08\pm0.05&  0.0       & -0.7\pm 1.3& -2.8\pm 1.1&  6.2\pm 1.2&  3.2\pm 0.8&  0.0       &  7.1\pm 1.6&0.17\pm0.02&0.33\pm0.02&  4.2\pm 0.7&  3.3\pm 0.8&  1.9\pm 1.0&   4.4\pm  0.6& \mathrm{Ab-spiral} \\        
 15&  ....     &  ....      &  3.9\pm 1.8& -2.1\pm 1.7&  5.7\pm 2.0&  5.8\pm 1.2&  0.0       & 10.2\pm 2.3&0.11\pm0.03&0.25\pm0.04&  3.8\pm 1.2&  3.9\pm 1.3&  6.0\pm 1.5&   4.1\pm  1.2& \mathrm{HDS_{red}}  \\	  
 16&1.80\pm0.05&  ....      &  0.8\pm 1.7& -2.4\pm 1.4&  5.8\pm 1.4&  2.1\pm 0.9&  0.0       &  5.9\pm 1.8&0.14\pm0.02&0.27\pm0.03&  6.3\pm 0.8&  3.3\pm 0.9&  3.8\pm 1.1&   1.7\pm  0.8& \mathrm{P} \\        
 17&1.31\pm0.03&  ....       &  9.5\pm 0.7&  8.0\pm 0.7&  3.0\pm 1.1&  4.8\pm 0.6& -1.0\pm 0.5&  3.6\pm 1.4&0.07\pm0.02&0.15\pm0.02&  3.5\pm 0.7&  4.0\pm 0.7&  3.1\pm 0.9&  -7.5\pm  0.9&\mathrm{ELG}\\   
 18&  ....     &  ....      & -1.1\pm 1.7& -3.8\pm 1.5&  5.8\pm 1.5&  2.5\pm 1.0&  0.0       &  9.0\pm 1.9&0.16\pm0.02&0.34\pm0.03&  6.9\pm 0.9&  4.5\pm 1.0&  1.2\pm 1.3&   2.5\pm  1.0& \mathrm{P} \\        
 19&1.98\pm0.05&  ....      & -0.6\pm 1.0&  1.5\pm 0.8&  6.1\pm 0.8&  2.4\pm 0.5&  0.0       &  4.9\pm 1.0&0.15\pm0.01&0.26\pm0.01&  5.0\pm 0.4&  3.0\pm 0.5&  7.4\pm 0.6&   0.5\pm  0.4& \mathrm{P} \\        
 20&1.66\pm0.04&  0.0       &  1.4\pm 3.0& -8.3\pm 2.7&  4.5\pm 2.9&  6.6\pm 1.7&  0.0       &  7.2\pm 3.4&0.13\pm0.04&0.30\pm0.05&  2.7\pm 1.7&  2.7\pm 1.8&  2.2\pm 2.1&   1.7\pm  1.6& \mathrm{Ab-spiral} \\        
 21&1.82\pm0.04&  0.0       & -1.2\pm 1.1& -3.1\pm 0.9&  4.9\pm 1.1&  2.5\pm 0.7&  0.0       &  7.5\pm 1.2&0.13\pm0.02&0.29\pm0.02&  5.0\pm 0.6&  3.4\pm 0.6&  2.8\pm 0.8&   3.4\pm  0.5& \mathrm{P} \\        
 22&1.17\pm0.05&  ....      &  4.6\pm 2.1&  1.7\pm 1.0&  8.7\pm 1.2& -3.0\pm 0.9& -2.5\pm 0.5&  7.1\pm 1.5&0.04\pm0.02&0.10\pm0.02&  4.7\pm 0.6&  2.5\pm 0.8&  7.2\pm 0.7&  -5.6\pm  0.7& \mathrm{ELG} \\	  
 23&1.67\pm0.04&  0.0       & -2.1\pm 0.9& -4.0\pm 0.8&  4.9\pm 0.9&  2.1\pm 0.6&  0.0       &  6.0\pm 1.1&0.13\pm0.01&0.26\pm0.02&  4.6\pm 0.5&  4.9\pm 0.6&  3.3\pm 0.7&   2.7\pm  0.5& \mathrm{P} \\        
            \noalign{\smallskip}	     		    
            \hline			    		    
         \end{array}
     $$ 
}
         \end{table}
\end{landscape}

\begin{landscape}
\addtocounter{table}{-1}
\begin{table}
          \caption[ ]{\footnotesize Continued.}
{\footnotesize
     $$ 
           \begin{array}{c c c c c c c c c c c c c c c c}
            \hline
            \noalign{\smallskip}
            \hline
            \noalign{\smallskip}
\mathrm{ID} &\mathrm{D_n(4000)} &\mathrm{[OII]} &\mathrm{H_{\delta A}} &
\mathrm{H_{\gamma A}} &\mathrm{Fe4531} &\mathrm{H_{\beta}} &\mathrm{[OIII]} 
&\mathrm{Fe5015} &\mathrm{Mg_1} &\mathrm{Mg_2} &\mathrm{Mg}b &
\mathrm{Fe5270} &\mathrm{Fe5335} &\mathrm{H_{\alpha}} & \mathrm{Spectral}\\
 &\AA &\AA &\AA &\AA &\AA &\AA &\AA &\AA &
\mathrm{mag} &\mathrm{mag} &\AA &\AA &\AA &\AA & \mathrm{class}\\
            \hline
            \noalign{\smallskip}   
 24&1.38\pm0.06&  ....      &  7.8\pm 1.2&  7.7\pm 1.1&  4.0\pm 1.5&  7.2\pm 0.8&  0.0       &  7.4\pm 1.8&-.02\pm0.02&0.11\pm0.03&  3.8\pm 0.9&  4.9\pm 1.1& -0.6\pm 1.5&   0.7\pm  1.1& \mathrm{HDS_{blue}} \\  
 25&1.84\pm0.11&  0.0       & -5.2\pm 1.9& -5.5\pm 1.3&  3.7\pm 1.3&  5.3\pm 0.7&  0.0       &  6.7\pm 1.4&0.15\pm0.02&0.30\pm0.02&  4.8\pm 0.6&  3.5\pm 0.7&  3.3\pm 0.9&  ....        & \mathrm{P} \\        
 26&1.70\pm0.03&  ....      & -1.6\pm 1.0& -2.8\pm 0.8&  5.2\pm 1.0&  2.5\pm 0.6&  0.0       &  6.0\pm 1.1&0.11\pm0.01&0.27\pm0.02&  4.6\pm 0.6&  2.5\pm 0.6&  4.2\pm 0.7&   2.9\pm  0.5& \mathrm{P} \\        
 27&  ....     &  ....      &  4.6\pm 2.2& -7.3\pm 2.2& 11.8\pm 1.8& -0.3\pm 1.3&  0.0       &  7.6\pm 2.5&0.18\pm0.03&0.38\pm0.04&  7.6\pm 1.1&  5.9\pm 1.3&  4.9\pm 1.5&   7.1\pm  1.0& \mathrm{HDS_{red}}\\              
 28&1.67\pm0.04&  0.0       & -2.0\pm 1.3& -2.8\pm 1.1&  4.9\pm 1.1&  5.1\pm 0.7&  0.0       &  7.4\pm 1.4&0.13\pm0.02&0.30\pm0.02&  4.8\pm 0.7&  2.8\pm 0.7&  2.8\pm 0.9&   ....       & \mathrm{P} \\         
 29&1.58\pm0.06&  0.0       &  2.6\pm 1.4&  3.9\pm 1.1&  5.6\pm 1.2&  5.8\pm 0.7&  0.0       &  5.4\pm 1.5&0.04\pm0.02&0.16\pm0.02&  3.4\pm 0.7&  4.9\pm 0.8&  3.2\pm 0.9&   3.6\pm  0.7& \mathrm{Ab-spiral} \\   	
 30&1.66\pm0.03&  0.0       & -0.8\pm 0.6& -4.1\pm 0.5&  4.3\pm 0.6&  2.0\pm 0.4&  0.0       &  5.2\pm 0.7&0.12\pm0.01&0.29\pm0.01&  4.9\pm 0.3&  3.1\pm 0.4&  2.1\pm 0.5&  -0.3\pm  0.4& \mathrm{P} \\         
 31&1.87\pm0.08&  0.0       & -1.5\pm 2.2& -1.5\pm 1.6& -1.9\pm 1.8&  4.1\pm 0.9&  0.0       &  6.9\pm 1.7&0.15\pm0.02&0.30\pm0.03&  6.2\pm 0.8&  5.5\pm 0.9&  2.5\pm 1.0&   1.2\pm  0.7& \mathrm{P} \\         
 32&1.83\pm0.04&  0.0       &  0.6\pm 0.7& -4.5\pm 0.7&  4.0\pm 0.7&  3.0\pm 0.5&  0.0       &  7.6\pm 0.9&0.15\pm0.01&0.29\pm0.01&  4.2\pm 0.4&  2.8\pm 0.5&  1.2\pm 0.6&   3.2\pm  0.4& \mathrm{P} \\         
 33&1.83\pm0.06&  0.0       &  0.8\pm 0.8& -3.6\pm 0.8&  3.4\pm 0.8&  2.7\pm 0.5&  0.0       &  5.6\pm 1.0&0.21\pm0.01&0.39\pm0.02&  6.4\pm 0.5&  3.5\pm 0.5&  3.0\pm 0.6&   2.3\pm  0.5& \mathrm{P} \\         
 34&1.58\pm0.10&  0.0       &  4.5\pm 1.5& -3.7\pm 1.3&  8.5\pm 1.3&  4.9\pm 0.7&  0.0       &  7.0\pm 1.5&0.09\pm0.02&0.32\pm0.02&  6.4\pm 0.7&  5.1\pm 0.8&  0.9\pm 1.0&   2.7\pm  0.6& \mathrm{HDS_{red}} \\       
 35&1.73\pm0.04&  0.0       & -2.1\pm 0.9& -3.5\pm 0.8&  5.6\pm 0.8&  3.6\pm 0.5&  0.0       &  8.3\pm 1.0&0.13\pm0.01&0.40\pm0.02&  7.1\pm 0.4&  4.4\pm 0.5&  4.7\pm 0.5&   ....       & \mathrm{P} \\         
 36&1.65\pm0.06&  ....      &  0.2\pm 2.3& -3.5\pm 1.7&  2.7\pm 1.9&  2.5\pm 1.1&  0.0       &  8.8\pm 2.1&0.14\pm0.03&0.32\pm0.03&  4.9\pm 1.0& 12.1\pm 1.0&  3.8\pm 1.3&   4.9\pm  0.9& \mathrm{P} \\         
 37&1.97\pm0.05&  0.0       & -1.2\pm 0.9& -5.1\pm 0.8&  4.8\pm 0.7&  1.4\pm 0.5&  0.0       &  4.0\pm 1.0&0.18\pm0.01&0.43\pm0.02&  6.5\pm 0.4&  2.8\pm 0.5&  2.4\pm 0.6&   2.4\pm  0.5& \mathrm{P} \\         
 38&1.95\pm0.07&  0.0       & -4.0\pm 1.3& -3.3\pm 1.0&  6.7\pm 0.9&  3.2\pm 0.5&  0.0       &  4.5\pm 1.0&0.21\pm0.01&0.39\pm0.02&  6.3\pm 0.5&  4.4\pm 0.5&  2.9\pm 0.7&  -0.4\pm  0.4& \mathrm{P} \\                 
 39&1.79\pm0.04&  0.0       & -1.4\pm 0.8& -4.0\pm 0.7&  4.7\pm 0.8&  3.7\pm 0.5&  0.0       &  7.1\pm 0.9&0.14\pm0.01&0.36\pm0.01&  6.4\pm 0.4&  3.5\pm 0.5&  3.1\pm 0.6&   2.3\pm  0.5& \mathrm{Ab-spiral} \\         
 40&1.56\pm0.10&  ....      & -3.6\pm 2.3&  1.8\pm 1.6&  3.7\pm 1.9&  2.0\pm 1.1&  0.0       &  8.8\pm 2.1&0.04\pm0.03&0.20\pm0.03&  4.1\pm 0.9&  4.0\pm 1.1&  5.3\pm 1.3&   2.1\pm  0.9& \mathrm{P} \\         
 41&1.85\pm0.07&  ....      &  2.3\pm 1.8& -2.2\pm 1.4&  9.0\pm 1.3&  2.2\pm 0.9&  0.0       & 10.4\pm 1.6&0.14\pm0.02&0.29\pm0.03&  4.3\pm 0.7&  2.4\pm 0.9&  3.4\pm 1.0&   4.3\pm  0.6& \mathrm{P} \\         
 42&1.56\pm0.10&  ....      & -0.8\pm 2.2&  3.9\pm 1.6&  0.9\pm 2.0&  3.6\pm 1.3&  0.0       &  1.3\pm 2.6&0.12\pm0.03&0.29\pm0.04&  5.6\pm 1.2&  5.4\pm 1.2&  2.3\pm 1.5&   1.9\pm  1.1& \mathrm{P} \\   
 43&1.19\pm0.03&  0.0       &  9.9\pm 0.3&  6.7\pm 0.3&  2.4\pm 0.5&  7.7\pm 0.3&  0.0       &  4.5\pm 0.6&0.04\pm0.01&0.13\pm0.01&  1.7\pm 0.3&  2.3\pm 0.4&  2.5\pm 0.5&   1.5\pm  0.4& \mathrm{HDS_{blue}} \\   
 44&1.93\pm0.05&  0.0       & -2.4\pm 1.4& -2.2\pm 1.1&  4.9\pm 1.3&  4.2\pm 0.7&  0.0       &  7.1\pm 1.6&0.15\pm0.02&0.38\pm0.02&  5.5\pm 0.7&  3.5\pm 0.8&  3.5\pm 0.9&  -1.9\pm  0.8& \mathrm{P} \\         
 45&1.98\pm0.08&  ....      & -2.9\pm 1.8& -3.7\pm 1.4&  4.7\pm 1.5&  3.6\pm 0.8&  0.0       &  3.6\pm 1.7&0.22\pm0.02&0.28\pm0.02&  7.2\pm 0.7&  6.1\pm 0.8&  4.2\pm 1.0&   2.3\pm  0.8& \mathrm{P} \\         
 46&1.83\pm0.06&  ....      & -2.6\pm 1.2& -4.7\pm 1.0&  6.9\pm 1.1&  3.2\pm 0.6&  0.0       &  9.0\pm 1.2&0.24\pm0.02&0.33\pm0.02&  1.5\pm 0.6&  2.0\pm 0.7&  3.0\pm 0.8&  -1.4\pm  0.6& \mathrm{P} \\        	
 47&1.70\pm0.04&  0.0       & -0.5\pm 0.8& -2.8\pm 0.7&  5.6\pm 0.8&  1.6\pm 0.5&  0.0       &  6.1\pm 1.0&0.17\pm0.01&0.35\pm0.02&  4.5\pm 0.5&  3.9\pm 0.5&  3.6\pm 0.6&   3.4\pm  0.5& \mathrm{P} \\		
 48&1.57\pm0.08&  0.0       & -1.3\pm 1.8& -2.1\pm 1.5&  4.1\pm 1.5&  2.5\pm 0.9&  0.0       &  9.4\pm 1.6&0.19\pm0.02&0.29\pm0.03&  6.1\pm 0.8&  2.1\pm 0.9&  2.9\pm 1.0&   3.4\pm  0.7& \mathrm{P} \\         
 49&1.82\pm0.03&  ....      & -0.9\pm 0.7& -1.6\pm 0.6&  6.0\pm 0.6&  2.6\pm 0.4&  0.0       &  6.0\pm 0.7&0.12\pm0.01&0.30\pm0.01&  4.6\pm 0.3&  3.2\pm 0.4&  3.2\pm 0.5&   ....       & \mathrm{P} \\         
 50&  ....     &  ....      & -1.5\pm 0.7& -5.7\pm 0.6&  4.0\pm 0.6&  2.4\pm 0.4&  0.0       &  5.8\pm 0.7&0.17\pm0.01&0.35\pm0.01&  4.3\pm 0.3&  3.6\pm 0.4&  2.9\pm 0.4&   3.7\pm  0.3& \mathrm{P} \\         
                \noalign{\smallskip}	     		    
            \hline			     		    
         \end{array}
     $$ 
}
         \end{table}
\end{landscape}

\begin{landscape}
\addtocounter{table}{-1}
\begin{table}
          \caption[ ]{\footnotesize Continued.}
{\footnotesize
     $$ 
           \begin{array}{c c c c c c c c c c c c c c c c c}
            \hline
            \noalign{\smallskip}
            \hline
            \noalign{\smallskip}
\mathrm{ID} &
\mathrm{D_n(4000)} &\mathrm{[OII]} &\mathrm{H_{\delta A}} &
\mathrm{H_{\gamma A}} &\mathrm{Fe4531} &\mathrm{H_{\beta}} &\mathrm{[OIII]} 
&\mathrm{Fe5015} &\mathrm{Mg_1} &\mathrm{Mg_2} &\mathrm{Mg}b &
\mathrm{Fe5270} &\mathrm{Fe5335} &\mathrm{H_{\alpha}} & \mathrm{Spectral}\\
 &\AA &\AA &\AA &\AA &\AA &\AA &\AA &\AA &
\mathrm{mag} &\mathrm{mag} &\AA &\AA &\AA &\AA & \mathrm{class}\\
            \hline
            \noalign{\smallskip}  
 51&1.65\pm0.04&  0.0       & -1.0\pm 0.7& -1.1\pm 0.6&  3.4\pm 0.7&  3.6\pm 0.4&  0.0       &  6.1\pm 0.8&0.12\pm0.01&0.29\pm0.01&  4.3\pm 0.4&  3.9\pm 0.5&  2.9\pm 0.6&   4.1\pm  0.4& \mathrm{P} \\         
 52&1.81\pm0.04&  0.0       & -1.8\pm 0.9& -3.3\pm 0.8&  3.8\pm 0.8&  2.4\pm 0.5&  0.0       &  5.8\pm 1.0&0.17\pm0.01&0.37\pm0.02&  5.7\pm 0.5&  2.5\pm 0.5&  2.8\pm 0.6&   2.3\pm  0.5& \mathrm{P} \\         
 53&1.51\pm0.09&  0.0       &  2.9\pm 3.3& -5.0\pm 2.3&  4.5\pm 2.2&  1.2\pm 1.2&  0.0       &  5.4\pm 2.3&0.15\pm0.03&0.34\pm0.04&  5.0\pm 1.0&  2.1\pm 1.2&  6.1\pm 1.3&   4.1\pm  0.9& \mathrm{P} \\       
 54&2.11\pm0.05&  0.0       & -1.2\pm 0.9& -2.3\pm 0.8&  4.1\pm 0.8&  3.4\pm 0.5&  0.0       &  7.8\pm 0.9&0.17\pm0.01&0.39\pm0.02&  5.7\pm 0.5&  4.3\pm 0.5&  2.4\pm 0.7&   2.3\pm  0.4& \mathrm{P} \\         
 55&1.77\pm0.04&  0.0       & -2.6\pm 0.6& -6.4\pm 0.6&  4.9\pm 0.6&  3.7\pm 0.4&  0.0       &  5.0\pm 0.7&0.17\pm0.01&0.41\pm0.01&  6.5\pm 0.3&  2.2\pm 0.4&  3.2\pm 0.5&   1.4\pm  0.4& \mathrm{P} \\         
 56&1.86\pm0.05&  0.0       & -2.1\pm 1.0& -4.3\pm 0.8&  2.6\pm 0.8&  3.3\pm 0.5&  0.0       &  4.9\pm 0.9&0.15\pm0.01&0.38\pm0.01&  6.4\pm 0.4&  2.9\pm 0.5&  2.6\pm 0.6&   0.7\pm  0.4& \mathrm{P} \\         
 57&  ....     &  ....      &  4.5\pm 1.3&  0.3\pm 1.2&  4.0\pm 1.4&  1.5\pm 0.8&  0.0       &  4.4\pm 1.6&0.08\pm0.02&0.18\pm0.02&  3.2\pm 0.8&  4.9\pm 0.8&  3.3\pm 1.0&   0.5\pm  0.7& \mathrm{HDS_{blue}} \\   
 58&1.66\pm0.03&  0.0       & -1.4\pm 0.8& -5.6\pm 0.7&  5.0\pm 0.8&  0.6\pm 0.5&  0.0       &  5.5\pm 1.0&0.17\pm0.01&0.39\pm0.02&  4.6\pm 0.5&  3.4\pm 0.5&  3.3\pm 0.7&   2.5\pm  0.5& \mathrm{P} \\         
 59&1.73\pm0.08&  0.0       &  6.4\pm 1.3& -2.9\pm 1.1&  6.4\pm 1.1& -0.4\pm 0.7&  0.0       &  8.9\pm 1.1&0.14\pm0.01&0.29\pm0.02&  5.9\pm 0.6&  3.7\pm 0.6&  2.3\pm 0.8&   1.9\pm  0.5& \mathrm{HDS_{red}} \\       
 60&1.70\pm0.06&  0.0       &  0.5\pm 1.3& -3.3\pm 1.0&  4.8\pm 1.0&  3.0\pm 0.6&  0.0       &  6.1\pm 1.2&0.15\pm0.01&0.39\pm0.02&  7.5\pm 0.5&  4.3\pm 0.6&  4.0\pm 0.7&   1.9\pm  0.5& \mathrm{P} \\         
 61&1.69\pm0.04&  0.0       & -3.5\pm 2.5& -2.5\pm 2.1&  4.8\pm 2.4&  2.3\pm 1.5&  0.0       &  7.7\pm 2.7&0.19\pm0.04&0.34\pm0.05&  5.5\pm 1.3&  2.6\pm 1.6&  4.4\pm 1.7&   2.0\pm  1.2& \mathrm{P} \\         
 62&1.09\pm0.01&-57.6\pm 1.8&  4.5\pm 1.1& -3.4\pm 1.2&  6.5\pm 1.3&-19.1\pm 1.2&-22.9\pm 0.8&-11.4\pm 1.9&0.06\pm0.02&0.05\pm0.02&  0.7\pm 0.9&  1.2\pm 1.0&  0.7\pm 1.1&-123.3\pm  1.6& \mathrm{?} \\ 
 63&  ....     &  ....      & -2.2\pm 1.1& -3.2\pm 1.0&  4.4\pm 0.9&  2.5\pm 0.5&  0.0       &  5.1\pm 1.1&0.16\pm0.01&0.37\pm0.02&  5.1\pm 0.5&  4.2\pm 0.6&  2.8\pm 0.7&   1.3\pm  0.5& \mathrm{P} \\         
 64&  ....     &  ....      &  3.2\pm 1.3& -0.1\pm 1.3&  4.4\pm 1.5&  0.6\pm 1.1&  0.0       &  4.8\pm 1.9&0.08\pm0.02&0.20\pm0.03&  2.8\pm 0.9&  4.6\pm 1.0&  2.4\pm 1.3& -10.6\pm  1.1& \mathrm{HDS_{blue}} \\   
 65&1.63\pm0.04&  0.0       & -3.2\pm 1.2& -3.5\pm 1.0&  6.3\pm 1.1&  3.3\pm 0.7&  0.0       &  7.8\pm 1.4&0.15\pm0.02&0.21\pm0.02&  6.9\pm 0.7&  4.5\pm 0.8&  0.8\pm 0.9&  -1.0\pm  0.8& \mathrm{Ab-spiral} \\         
 66&  ....     &  ....      & -1.0\pm 1.6& -2.1\pm 1.3&  4.4\pm 1.2&  1.8\pm 0.8&  0.0       &  7.0\pm 1.5&0.15\pm0.02&0.32\pm0.02&  5.3\pm 0.6&  3.6\pm 0.8&  3.1\pm 0.9&   2.1\pm  0.6& \mathrm{P} \\         
 67&1.54\pm0.10&  ....      &  1.2\pm 2.5&  4.4\pm 1.7& 10.5\pm 1.7&  4.5\pm 1.1&  0.0       & -0.7\pm 2.2&0.17\pm0.03&0.34\pm0.03&  2.9\pm 1.0&  2.7\pm 1.1&  2.1\pm 1.3&   ....       & \mathrm{P} \\         
 68&1.85\pm0.03&  0.0       & -2.3\pm 0.5& -4.3\pm 0.5&  4.6\pm 0.5&  4.2\pm 0.3&  0.0       &  6.5\pm 0.6&0.17\pm0.01&0.38\pm0.01&  5.8\pm 0.3&  3.5\pm 0.3&  3.1\pm 0.4&   1.5\pm  0.3& \mathrm{P} \\         
 69&2.01\pm0.06&  0.0       & -1.3\pm 1.3& -5.8\pm 1.1&  4.3\pm 1.0&  1.4\pm 0.7&  0.0       &  6.5\pm 1.2&0.17\pm0.01&0.40\pm0.02&  7.7\pm 0.5&  4.5\pm 0.6&  2.3\pm 0.7&   2.5\pm  0.5& \mathrm{P} \\         
 70&  ....     &  ....      &  1.2\pm 1.7& -3.1\pm 1.4&  2.9\pm 1.4&  0.0       &  0.0       &  5.1\pm 1.7&0.13\pm0.02&0.23\pm0.02&  3.8\pm 0.7&  2.5\pm 0.8&  3.7\pm 1.0&   1.7\pm  0.5& \mathrm{P} \\         
 71&1.98\pm0.04&  0.0       & -1.1\pm 0.9& -3.0\pm 0.7&  4.6\pm 0.8&  3.3\pm 0.5&  0.0       &  5.2\pm 0.9&0.18\pm0.01&0.36\pm0.02&  5.1\pm 0.4&  3.9\pm 0.5&  2.2\pm 0.6&   ....       & \mathrm{P} \\         
 72&1.82\pm0.11&  ....      &  4.9\pm 2.4&-11.1\pm 2.2&  4.4\pm 2.0&  5.1\pm 1.2&  0.0       &  2.4\pm 2.3&0.20\pm0.03&0.22\pm0.03&  5.5\pm 1.0&  6.0\pm 1.1&  4.3\pm 1.3&   3.6\pm  0.9& \mathrm{HDS_{red}} \\      
 73&1.87\pm0.04&  0.0       & -0.9\pm 0.8& -3.7\pm 0.6&  4.2\pm 0.7&  3.7\pm 0.4&  0.0       &  8.4\pm 0.9&0.16\pm0.01&0.39\pm0.01&  6.1\pm 0.4&  3.9\pm 0.4&  3.3\pm 0.6&   ....       & \mathrm{P} \\         
 74&1.71\pm0.04&  0.0       &  0.1\pm 0.4& -3.6\pm 0.4&  4.3\pm 0.5&  2.6\pm 0.3&  0.0       &  4.8\pm 0.7&0.14\pm0.01&0.22\pm0.01&  2.3\pm 0.3&  4.9\pm 0.4&  3.2\pm 0.5&   4.3\pm  0.3& \mathrm{P} \\         
 75&1.93\pm0.07& 0.0        &  3.0\pm 1.3& -5.6\pm 1.4&  2.0\pm 1.6&  3.3\pm 1.3&  0.0       &  2.6\pm 2.7&0.25\pm0.04&0.79\pm0.05& 16.7\pm 1.0& -1.7\pm 1.6&  0.6\pm 2.0&   ....       & \mathrm{P} \\         
 76&1.62\pm0.04&  0.0       & -1.8\pm 1.0& -4.0\pm 0.9&  2.7\pm 1.1&  6.0\pm 0.7&  0.0       &  3.1\pm 1.5&0.23\pm0.02&0.42\pm0.02&  4.4\pm 0.7&  3.1\pm 0.8&  5.2\pm 0.9&   ....       & \mathrm{P} \\         
 77&1.71\pm0.05&  0.0       & -2.7\pm 1.0& -6.1\pm 0.9&  5.1\pm 1.0&  1.9\pm 0.7&  0.0       &  5.5\pm 1.4&0.20\pm0.02&0.42\pm0.02&  5.6\pm 0.7&  1.4\pm 0.8&  2.5\pm 1.0&   ....       & \mathrm{P} \\         
            \noalign{\smallskip}	     		    
            \hline			    		    
         \end{array}
     $$ 
}
         \end{table}
\end{landscape}

\begin{landscape}
\addtocounter{table}{-1}
\begin{table}
          \caption[ ]{\footnotesize Continued.}
{\footnotesize
     $$ 
           \begin{array}{c c c c c c c c c c c c c c c c}
            \hline
            \noalign{\smallskip}
            \hline
            \noalign{\smallskip}
\mathrm{ID} &\mathrm{D_n(4000)} &\mathrm{[OII]} &\mathrm{H_{\delta A}} &
\mathrm{H_{\gamma A}} &\mathrm{Fe4531} &\mathrm{H_{\beta}} &\mathrm{[OIII]} 
&\mathrm{Fe5015} &\mathrm{Mg_1} &\mathrm{Mg_2} &\mathrm{Mg}b &
\mathrm{Fe5270} &\mathrm{Fe5335} &\mathrm{H_{\alpha}} & \mathrm{Spectral}\\
 & &\AA &\AA &\AA &\AA &\AA &\AA &\AA &\AA &
\mathrm{mag} &\mathrm{mag} &\AA &\AA &\AA & \mathrm{class}\\
            \hline
            \noalign{\smallskip}   
 78&  ....     &  ....       &  2.3\pm 2.9&  0.5\pm 3.1&  3.8\pm 3.5&  4.5\pm 2.2& 0.0       & 14.5\pm 5.9&0.02\pm0.08&0.15\pm0.10&  4.8\pm 3.3& 11.0\pm 3.2& -6.9\pm 6.1&   ....       & \mathrm{P} \\         
 79&  ....     &  ....      &  5.1\pm 0.6&  5.0\pm 0.5&  3.1\pm 0.7&  4.5\pm 0.4&  0.0       &  6.6\pm 0.8&0.08\pm0.01&0.19\pm0.01&  3.2\pm 0.4&  3.8\pm 0.5&  3.9\pm 0.6&   2.5\pm  0.5& \mathrm{HDS_{blue}} \\   
 80&  ....     &  ....      &  0.3\pm 1.6& -4.7\pm 1.6&  5.2\pm 1.8&  4.4\pm 1.1&  0.0       & 12.1\pm 2.2&0.10\pm0.03&0.38\pm0.04&  6.1\pm 1.1&  0.6\pm 1.4&  5.2\pm 1.4&  -3.6\pm  1.4& \mathrm{P} \\         
 81&1.31\pm0.05& -6.1\pm 1.7&  5.1\pm 1.3&  3.5\pm 1.3&  4.8\pm 1.6&  1.6\pm 1.1& -0.2\pm 0.8&  0.5\pm 2.3&0.07\pm0.03&0.23\pm0.04&  0.9\pm 1.2&  6.0\pm 1.1&  3.0\pm 1.4&  -4.1\pm  1.3& \mathrm{ELG} \\       
 82&  ....     &  ....      & -2.5\pm 1.3& -4.7\pm 1.1&  4.6\pm 1.0&  2.9\pm 0.6&  0.0       &  7.6\pm 1.2&0.14\pm0.02&0.35\pm0.02&  5.5\pm 0.6&  3.1\pm 0.6&  2.6\pm 0.7&   2.2\pm  0.5& \mathrm{P} \\            
 83&  ....     &  ....      &  3.3\pm 1.6& -3.9\pm 1.7& -1.9\pm 2.3&  2.9\pm 1.5&  0.0       &  5.4\pm 3.6&0.05\pm0.05&0.61\pm0.07& 11.8\pm 1.5&-11.8\pm 2.5&  5.1\pm 2.5&   ....       & \mathrm{HDS_{red}} \\         
 84&1.35\pm0.04& -6.5\pm 1.4&  9.0\pm 0.9&  2.4\pm 0.9&  0.7\pm 1.3&  3.7\pm 0.9& -1.1\pm 0.7&  9.3\pm 1.8&0.02\pm0.02&0.19\pm0.03&  3.9\pm 0.9&  0.0\pm 1.1&  2.1\pm 1.3&  -6.0\pm  1.5& \mathrm{ELG} \\          
 85&1.77\pm0.07&  0.0       &  3.0\pm 1.5&  0.3\pm 1.5&  0.5\pm 1.9&  7.1\pm 1.2&  0.0       & 10.7\pm 2.6&0.18\pm0.04&0.52\pm0.05&  7.2\pm 1.3&  0.8\pm 1.6&  5.5\pm 1.9&  -4.1\pm  2.8& \mathrm{P} \\            
 86&1.98\pm0.04&  ....      & -2.1\pm 0.7& -4.0\pm 0.6&  3.8\pm 0.6&  2.1\pm 0.4&  0.0       &  5.5\pm 0.7&0.16\pm0.01&0.35\pm0.01&  4.5\pm 0.3&  3.2\pm 0.4&  2.4\pm 0.5&   1.3\pm  0.3& \mathrm{P} \\            
 87&1.68\pm0.05&  0.0       & -1.3\pm 0.8& -3.1\pm 0.7&  2.7\pm 0.8&  2.2\pm 0.5&  0.0       & 10.5\pm 0.9&0.20\pm0.01&0.33\pm0.01&  4.7\pm 0.5&  3.5\pm 0.5&  0.6\pm 0.6&   0.0\pm  0.5& \mathrm{P} \\      	     
 88&1.60\pm0.07&  0.0       &  0.2\pm 1.8& -5.4\pm 1.7&  5.0\pm 1.9& -0.4\pm 1.4&  0.0       &  1.6\pm 2.8&0.04\pm0.03&0.17\pm0.04&  2.2\pm 1.5&  6.4\pm 1.5&  4.6\pm 1.9&   6.3\pm  2.0& \mathrm{P} \\            
 89&1.75\pm0.05&  0.0       & -0.8\pm 1.3& -1.6\pm 1.2&  4.8\pm 1.4&  4.3\pm 0.8&  0.0       &  6.5\pm 1.6&0.19\pm0.02&0.31\pm0.03&  4.4\pm 0.8&  4.5\pm 0.9&  1.9\pm 1.1&   0.6\pm  0.9& \mathrm{P} \\            
 90&  ....     &  ....      &  1.7\pm 3.1& -3.5\pm 2.5&  4.8\pm 2.3&  3.7\pm 1.4&  0.0       &  4.7\pm 2.6&0.11\pm0.03&0.25\pm0.04&  5.2\pm 1.1&  2.9\pm 1.3&  3.2\pm 1.5&   2.3\pm  0.8& \mathrm{P} \\            
 91&  ....     &  ....      & -4.1\pm 1.1& -4.7\pm 1.0&  4.1\pm 1.0&  2.9\pm 0.6&  0.0       &  5.1\pm 1.2&0.14\pm0.01&0.37\pm0.02&  6.3\pm 0.5&  4.0\pm 0.6&  3.8\pm 0.8&   1.9\pm  0.6& \mathrm{P} \\  	     
 92&1.35\pm0.04& -7.8\pm 1.9&  5.5\pm 1.2&  4.9\pm 1.2&  3.4\pm 1.5&  0.9\pm 1.0& -0.2\pm 0.7&  3.8\pm 2.1&0.10\pm0.03&0.22\pm0.03&  2.7\pm 1.1&  2.8\pm 1.2&  5.2\pm 1.3& -16.2\pm  1.8& \mathrm{ELG} \\	     
 93&  ....     &  ....      & -1.8\pm 0.7& -6.2\pm 0.6&  5.0\pm 0.7& -0.1\pm 0.4&  0.0       &  4.5\pm 0.8&0.20\pm0.01&0.38\pm0.01&  5.2\pm 0.4&  3.4\pm 0.4&  3.1\pm 0.5&   2.3\pm  0.3& \mathrm{P} \\            
 94&1.61\pm0.05&  ....      &  2.7\pm 0.9& -0.8\pm 0.9&  5.5\pm 1.0&  4.5\pm 0.6&  0.0       &  7.7\pm 1.2&0.11\pm0.02&0.36\pm0.02&  5.2\pm 0.6&  0.9\pm 0.7&  1.6\pm 0.9&  -0.4\pm  0.8& \mathrm{Ab-spiral} \\             
 95&  ....     &  ....      & -1.0\pm 0.9& -3.8\pm 0.8&  4.7\pm 0.8&  3.2\pm 0.5&  0.0       &  9.6\pm 0.9&0.19\pm0.01&0.37\pm0.01&  6.0\pm 0.4&  4.3\pm 0.5&  4.0\pm 0.6&   0.8\pm  0.4& \mathrm{P} \\             
 96&  ....     &  ....      & -1.5\pm 0.9& -4.6\pm 0.8&  3.5\pm 0.9&  4.6\pm 0.6&  0.0       &  3.9\pm 1.2&0.16\pm0.02&0.38\pm0.02&  6.3\pm 0.6&  3.1\pm 0.7&  3.3\pm 0.8&   1.4\pm  0.7& \mathrm{P} \\             
 97&1.55\pm0.08&  ....      &  0.2\pm 2.3& -3.0\pm 2.2&  3.3\pm 2.4&  2.5\pm 1.5&  0.0       & 13.2\pm 2.6&0.18\pm0.04&0.26\pm0.04&  4.5\pm 1.4& 11.7\pm 1.3& 10.4\pm 1.4&   4.9\pm  0.8& \mathrm{P} \\             
 98&1.82\pm0.05&  ....      & -2.7\pm 0.8& -6.8\pm 1.4&  4.4\pm 1.4&  2.2\pm 0.8&  0.0       & 10.2\pm 1.7&0.16\pm0.02&0.31\pm0.03&  1.5\pm 1.0&  5.7\pm 1.1&  3.0\pm 1.3&  -4.0\pm  1.9& \mathrm{P} \\             
 99&1.76\pm0.04&  ....      & -0.4\pm 0.9& -6.6\pm 0.8&  3.8\pm 0.8&  3.7\pm 0.5&  0.0       &  4.4\pm 1.0&0.11\pm0.01&0.30\pm0.02&  4.2\pm 0.5&  3.6\pm 0.5&  3.2\pm 0.6&   2.5\pm  0.5& \mathrm{P} \\             
100&  ....     &  ....      &  3.0\pm 1.6&-10.1\pm 1.8& -0.9\pm 2.2& 10.0\pm 1.4&  0.0       &  9.5\pm 3.8&0.02\pm0.04&0.21\pm0.05&  0.9\pm 1.8&  6.4\pm 1.8&  3.0\pm 2.2&  -3.4\pm  1.8& \mathrm{Ab-spirals} \\	     
101&  ....     &  ....      &  0.3\pm 2.6& -4.4\pm 2.1&  3.4\pm 2.1&  6.7\pm 1.2&  0.0       & 14.6\pm 2.6&0.18\pm0.04&0.34\pm0.05& 10.2\pm 1.2&  6.4\pm 1.4& -3.3\pm 2.2&   1.8\pm  1.6& \mathrm{P} \\	     
102&  ....     &  ....      &  4.5\pm 1.0& -1.0\pm 1.0&  4.0\pm 1.3&  1.8\pm 0.9& -1.6\pm 0.7&  8.8\pm 1.9&0.03\pm0.02&0.11\pm0.03&  3.6\pm 1.0& -0.6\pm 1.2&  2.3\pm 1.3&  -6.9\pm  1.3& \mathrm{ELG}\\                    
            \noalign{\smallskip}	     		   
            \hline
            \noalign{\smallskip}
            \hline
         \end{array}
     $$ 
}
         \end{table}
\end{landscape}

\begin{acknowledgements}
We are grateful to the referee , T. Goto, for his very accurate work
on the paper.  We are grateful the to Stephan Charlot and Gustavo
Bruzual who provided us with the galaxy models. We thank Andrea
Biviano for useful discussions. A. M. thanks Massimo Capaccioli for
useful discussions and for the hospitality at the INAF-Osservatorio
Astronomico di Capodimonte, and Francesca Matteucci for support during
this work.

\end{acknowledgements}

\end{document}